\documentclass[sigconf]{acmart}
\AtBeginDocument{%
 \providecommand\BibTeX{{%
 \normalfont B\kern-0.5em{\scshape i\kern-0.25em b}\kern-0.8em\TeX}}}

\copyrightyear{2025}
\acmYear{2025}
\setcopyright{cc}
\setcctype{by-nc}
\acmConference[CHI '25]{CHI Conference on Human Factors in Computing Systems}{April 26-May 1, 2025}{Yokohama, Japan}
\acmBooktitle{CHI Conference on Human Factors in Computing Systems (CHI '25), April 26-May 1, 2025, Yokohama, Japan}\acmDOI{10.1145/3706598.3713849}
\acmISBN{979-8-4007-1394-1/25/04}

\usepackage{multirow}
\usepackage{subcaption}
\usepackage{balance}       
\usepackage{graphics}      
\usepackage{hyperref}
\usepackage{color}
\usepackage{textcomp}
\usepackage{subcaption}
\usepackage{xcolor}
\usepackage{makecell}
\usepackage{multirow}
\usepackage{enumitem}
\usepackage{amsmath}
\usepackage{stfloats}
\usepackage{graphicx}
\usepackage{amsthm}
\usepackage{listings}
\usepackage{xspace}
\usepackage[linesnumbered]{algorithm2e}

\newcolumntype{L}[1]{>{\raggedright\let\newline\\\arraybackslash\hspace{0pt}}m{#1}}
\newcolumntype{C}[1]{>{\centering\let\newline\\\arraybackslash\hspace{0pt}}m{#1}}
\newcolumntype{R}[1]{>{\raggedleft\let\newline\\\arraybackslash\hspace{0pt}}m{#1}}

\makeatletter
\def\thickhline{%
  \noalign{\ifnum0=`}\fi\hrule \@height \thickarrayrulewidth \futurelet
   \reserved@a\@xthickhline}
\def\@xthickhline{\ifx\reserved@a\thickhline
               \vskip\doublerulesep
               \vskip-\thickarrayrulewidth
             \fi
      \ifnum0=`{\fi}}
\makeatother

\makeatletter
\def\thickhlinespace{%
  \addlinespace[1ex]
  \noalign{\ifnum0=`}\fi\hrule \@height \thickarrayrulewidth \futurelet
   \reserved@a\@xthickhline
   \addlinespace[1ex]
   }
\def\@xthickhlinespace{\ifx\reserved@a\thickhline
               \vskip\doublerulesep
               \vskip-\thickarrayrulewidth
             \fi
      \ifnum0=`{\fi}}
\makeatother

\newlength{\thickarrayrulewidth}
\newlength\Origarrayrulewidth

\definecolor{downredcolor}{HTML}{e31a1c}
\definecolor{upgreencolor}{HTML}{33a02c}

\definecolor{DarkGreen}{HTML}{5DAC81}

\usepackage{enumitem}
\setitemize{noitemsep,topsep=0pt,parsep=0pt,partopsep=0pt}

\usepackage[defaultcolor=blue,commandnameprefix=always]{changes}

\begin{document}


\title[Ego vs. Exo and Active vs. Passive in Immersive Storytelling]{Ego vs. Exo and Active vs. Passive: \\ Investigating the Effects of Viewpoint and Navigation on Spatial Immersion and Understanding in Immersive Storytelling}

\author{Tao Lu}
\affiliation{%
  \institution{Georgia Institute of Technology}
  \city{Atlanta, GA}
  \country{USA}
}
\email{luttul@gatech.edu}

\author{Qian Zhu}
\affiliation{%
  \institution{Renmin University of China}
  \city{Beijing}
  \country{China}
}
\email{qzhual@connect.ust.hk}

\author{Tiffany Ma}
\affiliation{%
  \institution{Georgia Institute of Technology}
  \city{Atlanta, GA}
  \country{USA}
}
\email{tiffany.ma@duke.edu}

\author{Wong Kam-Kwai}
\affiliation{%
  \institution{The Hong Kong University of Science and Technology}
  \city{Hong Kong}
  \country{China}
}
\email{kkwongar@connect.ust.hk}

\author{Anlan Xie}
\affiliation{%
  \institution{Georgia Institute of Technology}
  \city{Atlanta, GA}
  \country{USA}
}
\email{anlan.xie@gmail.com}

\author{Alex Endert}
\affiliation{%
  \institution{Georgia Institute of Technology}
  \city{Atlanta, GA}
  \country{USA}
}
\email{endert@gatech.edu}

\author{Yalong Yang}
\affiliation{%
  \institution{Georgia Institute of Technology}
  \city{Atlanta, GA}
  \country{USA}
}
\email{yalong.yang@gatech.edu}

\renewcommand{\shortauthors}{Lu et al.}



\begin{abstract}
Visual storytelling combines visuals and narratives to communicate important insights. While web-based visual storytelling is well-established, leveraging the next generation of digital technologies for visual storytelling, specifically immersive technologies, remains underexplored.
We investigated the impact of the story viewpoint (from the audience's perspective) and navigation (when progressing through the story) on spatial immersion and understanding. First, we collected web-based 3D stories and elicited design considerations from three VR developers. We then adapted four selected web-based stories to an immersive format. Finally, we conducted a user study (N=24) to examine egocentric and exocentric viewpoints, active and passive navigation, and the combinations they form.
Our results indicated significantly higher preferences for egocentric+active (higher agency and engagement) and exocentric+passive (higher focus on content). We also found a marginal significance of viewpoints on story understanding and a strong significance of navigation on spatial immersion.
\end{abstract}


\settopmatter{printfolios=true}
\begin{teaserfigure}
  \includegraphics[width=\textwidth]{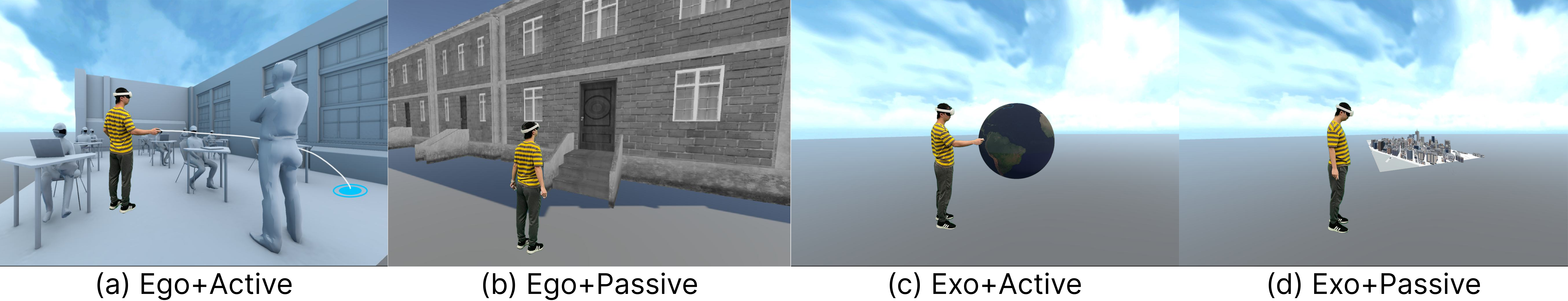} 
  \caption{We investigated the effects of different combinations of viewpoints and navigation in immersive storytelling, as demonstrated in four implemented scenarios: (a), (b), (c), and (d). In (a) and (b), the audience is fully immersed in the story scene from an egocentric perspective. Conversely, in scenarios (c) and (d), the audience maintains an exocentric perspective, remaining independent of the story. In (a) and (c), the audience navigates the story manually and actively, following specific instructions. In contrast, (b) and (d) allow audience clicking a controller button, which triggers the story and allows navigation to proceed automatically.}
  \Description{Four Combinations of Viewpoints and Navigations. The first scene (leftmost) depicts the person standing in a classroom setting with several figures seated around tables. The figures appear to be rendered in a basic gray, suggesting that they are part of a virtual environment. The second scene shows the person standing outside of what appears to be a virtual building, interacting with a door or window. The environment has a simplistic, grayish brick design. The third scene presents the person holding a 3D model of the Earth, as though they are interacting with a virtual globe. The final scene (rightmost) shows the person seemingly holding or interacting with a floating model of a cityscape, with buildings arranged on a platform that hovers in the virtual space.}
  \label{fig:teaser}
  \vspace{1em}
\end{teaserfigure}
\maketitle


\section{Introduction}
Visual storytelling uses the combination of visual elements and narratives to communicate complex ideas (e.g., data, facts, and opinions) in a way that is more engaging and descriptive than traditional text-only stories~\cite{kulkarni2023innovating}. 
This approach has been widely used in various fields, including journalism~\cite{nyt, wapo, guardian}, education~\cite{Williams2019Attending, Baharuddin2023The}, commercial activities~\cite{Moin2020Storytelling, Zhou2005VISUAL}, and scientific communication~\cite{Ma2012Scientific}. It plays a crucial role in enhancing the readers' understanding of content, as well as increasing their emotional engagement and connection to the story.

The evolution of digital technologies has greatly expanded the possibilities of visual storytelling, offering new genres and more diverse ways to deliver content~\cite{segel2010narrative}. 
The introduction of digital displays brought dynamic and interactive visual content, changing how readers interact and engage with stories~\cite{Yu2018A, Greussing2020Learning}. 
The rise of mobile devices further transformed access and interaction modalities, enabling readers to experience stories anywhere with new input modalities such as touch-based interactions~\cite{Lima2020Sketch-Based,Sheremetieva2022Touch}.
As digital technologies continue to evolve, the pressing question now is: "What is next for visual storytelling, and how can it offer novel experiences?"

As immersive technologies rapidly develop, they are becoming increasingly accessible and affordable, positioning them as promising candidates for the next generation of major public digital platforms. Immersive devices (e.g., Virtual Reality (VR) headsets) can render 2D and 3D information in the surrounding space or even replace the users' entire visual field, offering unprecedented opportunities for visual storytelling~\cite{kraus2021value, isenberg2018immersive}.
A prominent use of immersive technology for storytelling is 360-degree videos~\cite{google_2016_beyond, sportsillustrated_2017_chapter}. However, these videos often limit the readers' navigation, providing little control over story progression or detail exploration due to their fully author-driven nature.
Beyond static videos, major news outlets like The New York Times and The Washington Post have embraced interactive formats in their web-based articles to present public events (e.g., COVID-19 and elections) more dynamically.
They also leverage immersive technologies for more interactive experiences, such as 
mobile AR stories created by The New York Times~\cite{thenewyorktimes_2022_showcasing}. 
However, these narratives tend to rely on basic touchscreen inputs and offer limited immersion. They are often confined to a single scene without transitions between multiple narrative elements.
In this work, we focus on improving immersive storytelling specifically for public journalism.
We seek to extend the design of these journalistic stories into immersive environments, where interactivity and complete storylines can foster deeper engagement and understanding.

Our goal is to leverage the unique display and interaction affordances of immersive technologies in storytelling. 
A major advantage of these technologies is their ability to render 3D spatial information at its original scale~\cite{isenberg2018immersive}. 
Building on this characteristic, we investigated how people visually perceive 3D spatial representations (i.e., ego- vs. exocentric viewpoints) and how they can interactively navigate in the 3D space (i.e., active vs. passive navigation) as an initial exploration. Specifically, readers can experience an immersive story from an \textbf{exocentric} (exo) viewpoint to gain an overview of a 3D scene or from an \textbf{egocentric} (ego) viewpoint to fully immerse themselves with life-sized artifacts. Additionally, readers can either \textbf{actively} move in space and interact with digital artifacts with greater embodiment and agency or \textbf{passively} follow predefined transitions.
We anticipate that these two factors (i.e., viewpoint and navigation) will have a significant effect on user-perceived spatial immersion and understanding in immersive storytelling. 
\textbf{\textit{Our research goal is to understand the advantages and limitations of these design choices in spatial immersion and understanding in immersive VR stories}}.

\begin{figure*}[t]
\centering
  \includegraphics[width=0.6\linewidth]{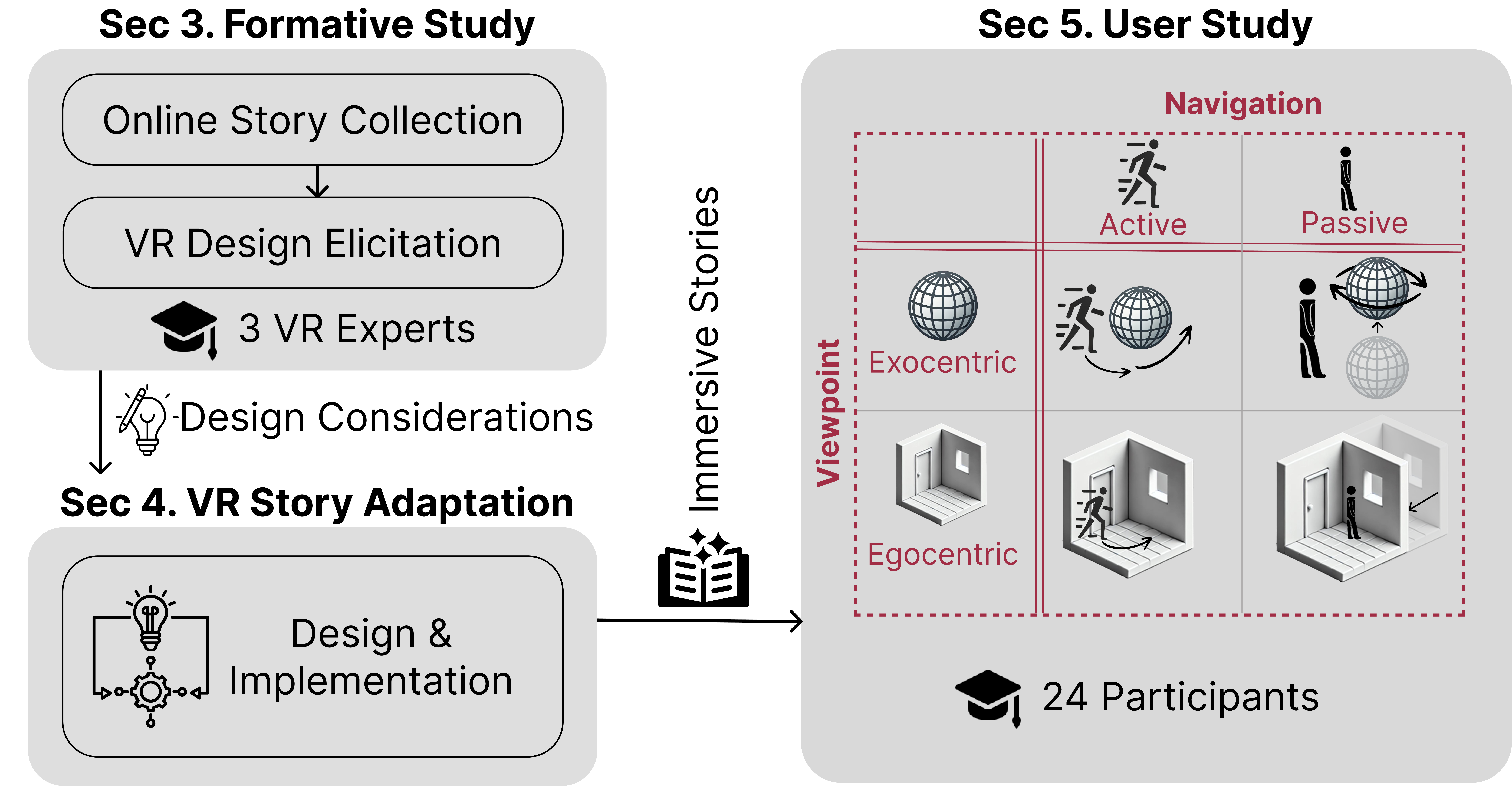}
  \caption{Overall Procedure of This Work. We started by a formative study analyzing the web-based stories in terms of their viewpoint and navigation designs. It elicited some design considerations, which we used to design and implemented four story cases. We finally investigated the effects of viewpoints and navigations on 24 participants.}
  \label{fig:procedure}
  \Description{A research workflow diagram illustrating the process of a VR storytelling study. The left side details the Formative Study (Section 3), starting with web-based story collection and VR design elicitation from 3 experts to identify design considerations. These considerations are applied to the VR Story Adaptation (Section 4), involving design and implementation of immersive stories. The right side describes the User Study (Section 5), which evaluates different combinations of viewpoint (egocentric vs. exocentric) and navigation (active vs. passive) with 24 participants.}
\end{figure*}

To study the effects of viewpoints and navigation in immersive storytelling, we needed to develop immersive stories that were controlled for these factors. 
While there are many web-based stories for public journalism, immersive versions are still rare. Since we aimed to focus on viewpoint and navigation rather than the story content itself, we leveraged existing web-based stories and created versions in immersive environments to cover the design variations we needed for the study.
We collected web-based stories from major news outlets, particularly those with rich 3D spatial content, such as a New York Times story about how COVID-19 spreads in a classroom~\cite{covid}. 
{These visual stories are in single scene and follow a linear structure. We consider this atomicity as a starting point for studying our intended effects. Meanwhile, this basic structure serves as the foundational unit for more complex stories.}
We then conducted a formative study with three VR experts to elicit potential designs of 117 selected stories in an immersive format, with an intended focus on viewpoint and navigation.
Based on the design considerations from this formative study, we adapted four representative stories into immersive formats. Finally, we conducted a user study with 24 participants to compare the 2$\times2$ combinations (ego vs. exo viewpoint, and active vs. passive navigation).
Our paper structure follows the steps we took to prepare and conduct the study, as illustrated in \autoref{fig:procedure}.

We found that viewpoints and navigations showed more significant impact on the story understanding and spatial immersion, respectively. Exo demonstrated benefits on story content comprehension and ego performed better on the spatial information memorability. Active navigation created a higher level of presence with a cost of more perceived workload, but it is acceptable in a ego viewpoint. Overall, participants predominantly preferred Ego$+$Active or Exo$+$Passive, depending on their priority on contents or experiences.

In summary, our primary contributions are:
\begin{itemize}
    \item {Four design considerations elicited for adapting web-based visual stories to an immersive format.}
    \item A user study empirically investigates the effects of viewpoint and navigation on perceived spatial immersion and understanding {across four adapted immersive story instances}.
\end{itemize}
\section{Related Works}
\noindent\textbf{Visual Storytelling.}
Visual storytelling is widely used in journalism to help audiences understand data-backed facts through static or dynamic visual elements, such as images, videos, and animations.
Segal and Heer~\cite{segel2010narrative} were among the first to summarize emerging narrative visualization techniques in web-based storytelling, discussing design variations across different genres.
Hullman et al.~\cite{hullman2011visualization} studied how information prioritization affects user interpretation. Kosara and Mackinlay~\cite{kosara2013storytelling} further highlighted the potential of storytelling in visualizations, emphasizing its promising future.
Subsequent research explored design considerations in visual storytelling, proposing guidelines to improve content representation and interaction. For instance, Morais et al.\cite{morais2020showing} proposed a design framework for anthropographic data stories, while Dasu et al.~\cite{dasu2023character} examined the use of characters in data stories to enhance enjoyment and persuasion. Yang et al.~\cite{yang2021design} also explored the application of narrative structures in data stories. These studies connect design practices with human cognition theories, offering valuable insights for visual storytelling design.

Beyond narrative structure, audience engagement and interaction are critical to the story’s impact, persuasion, and memorability.
With advancements in technology, visual storytelling has expanded to include videos, games, and VR/AR platforms~\cite{zhao2023stories,mendez2025immersive,zhou2023data}, offering audiences more interactive experiences than static texts and images alone. VR/AR, in particular, immerses users in virtual spaces, enhancing spatial coherence and unlocking new design possibilities.
This paper focuses on two key elements in VR/AR and visual storytelling: viewpoints and navigation, which shape user perspective and agency, directly influencing immersion, comprehension, and emotional engagement. We aim to identify effective design practices for integrating these aspects into immersive storytelling.

\noindent\textbf{Immersive Experiences.}
The goals of visual storytelling are multifaceted~\cite{amini2018evaluating}. For authors, common objectives include (a) communicating facts and insights, and (b) persuading audiences of the opinions conveyed through the story. For audiences, the primary goal is to gain new knowledge, such as news, insights, and perspectives, while entertainment also plays a significant role, as stories may be consumed for enjoyment.

While 2D screens are effective for visual storytelling, immersive technologies like VR/AR extend stories into 3D, closing the gap between audiences and narratives by placing them in the same space. Immersive experiences offer several advantages.
First, they enhance presence and immersion~\cite{servotte2020vr}. VR/AR can simulate real-world scenarios~\cite{zhao2023leanon} with real-time rendering and embodied interactions~\cite{huang2023embodied,zhu2024compositingvis,in2023table}, or even go beyond reality (e.g., ground-level scaling~\cite{abtahi2019m}).
Second, they improve understanding and retention~\cite{Motejlek2023The, Yildirim2019The}. VR/AR is frequently used in scientific visualizations~\cite{marriott2018immersive}, helping users extract and communicate insights more effectively.
Third, they enhance emotional communication~\cite{kandaurova2019effects}, narrowing the emotional distance between authors and audiences and fostering resonance and reflection on the story's themes~\cite{bahng2020reflexive}.

These benefits align with the core goals of visual storytelling. Therefore, this paper investigates the impact of immersive environments on understanding and spatial immersion.

\noindent\textbf{Immersive Storytelling.}
Immersive storytelling, a broader topic of integrating immersive experiences to different storytellings, has been studied in various other contexts. {In AR, Shin et al.~\cite{shin2019any} studied the two factors: room size and furniture density on a AR crime-solving game, and they suggested more context-awareness such as using virtual objects as substitutes in large and empty rooms. A later survey~\cite{shin2024investigating} furthered this idea, proposing to balance virtual and real experiences through situated user interactions. In VR,} Lee et al. leveraged VR to materialize abstract measure and units so that audiences could connect with their natural experiences~\cite{lee2020data}. Hall et al. focused on synchronous immersive data presentation and communication~\cite{hall2022augmented}. {VR storytelling is particularly relevant in cinematic.} Collen et al. investigated the cinematic techniques used in immersive narrative visualizations of 3D spatial information~\cite{conlen2023cinematic}. {Practically, Wu and Karwas~\cite{wu2024metamorphosis} developed a technique that allows free viewpoint change and with time-based interactivity.} Other immersive storytelling works focusing on specific applications, such as situated awareness of social problems~\cite{assor2024augmented, zhu2024reader} or emotional enjoyment of visualizations~\cite{romat2020dear}. These works contribute to the design of immersive storytelling by highlighting the benefits of immersive technologies we should leverage in specific topics or contexts, but none explored the representation and interaction design on visual jouralism (e.g. scrollytelling) in immersive environment. 

There are some studies focusing on how traditional storytellings on 2D screens can be transported to VR/AR, which are primarily based on the video format ~\cite{yang2023understanding, Zollmann2020CasualVRVideos, Nash2018Virtually}. The adaptation of interactive visual stories, a popular form of visual storytelling, are still underexplored. 

\noindent\textbf{Viewpoint and Navigation in Immersive Environments.}
In traditional digital media (e.g., computer monitors or mobile devices), content is typically viewed from an exocentric (third-person) viewpoint, offering a comprehensive overview of spatial relationships and layout. Immersive technology, while capable of maintaining an exocentric view, uniquely allows for a 360-degree stereoscopic experience from an egocentric (first-person) viewpoint. Despite this capability, the egocentric view is not widely adopted due to mixed findings on its effectiveness. 
Yang et al.~\cite{yang2018maps} found that exocentric views significantly outperformed egocentric views for geographic analysis tasks. Similarly, Kraus et al.~\cite{kraus2019impact} and Yang et al.~\cite{yang2020embodied} noted limitations in egocentric 3D scatterplot visualizations, such as lack of overview and targets being out of view. 
However, the egocentric view has shown promise in multi-view management within immersive environments~\cite{in2024evaluating,davidson2022exploring,satriadi2020maps,luo2022should}, and studies have demonstrated improved memory and knowledge retention in egocentric settings~\cite{krokos2019virtual,yang2020virtual}. 
{Notably, Hoppe et al.~\cite{hoppe2022there} argued for a perspective continuum in immersive environments, finding no difference in users' perceived workload, sense of presence, and engagement between egocentric and exocentric viewpoints in a VR combat game they developed. However, it remains questionable whether this perspective continuum is applicable in immersive storytelling, as storytelling involves different perception and cognitive processes compared to games.} 
Given these mixed results, our research explores the advantages and drawbacks of both exocentric and egocentric viewpoints in the context of visual storytelling.

{Navigation in immersive environments refers to the process of moving to the target location within 3D virtual environments. It is a crucial interaction for exploring 3D spaces in these settings. In the context of immersive storytelling, navigation plays a vital role in allowing users to visually explore different parts of the story and advance the narrative (e.g., a story progresses to the next stage once the viewer reaches a specific location).
Significant efforts have been devoted to developing various navigation techniques, including gesture design~\cite{tursunov2024creating}, utilizing different input modalities~\cite{swidrak2024beyond}, leveraging perception and space manipulation~\cite{dong2021tailored}, and experimenting with different scales~\cite{mirhosseini2019exploration, abtahi2019m}. 
Luca et al.~\cite{di2021locomotion} conducted an extensive survey of these techniques and compiled a Locomotion Vault. However, some of these innovative techniques require additional hardware, while others demand a steep learning curve.
As a result, the mainstream and default navigation techniques adopted by major platforms remain natural walking and teleportation. Considering the accessibility aspect of storytelling, we decided to focus on using mainstream navigation techniques for a broader user group. In many immersive applications, users need to actively navigate the 3D space to complete tasks. However, in storytelling, we have the option to let users passively follow a predefined path without explicit navigation actions (such as walking or teleportation).
While active navigation provides more interactivity, it carries the risk of disorienting the user. In contrast, passive navigation maintains narrative control but may reduce the sense of agency and induce motion sickness. Deciding which method to use is a unique and important consideration for immersive storytelling.
In relation to this topic, Lages and Bowman~\cite{lages2018move} compared physical walking (where the user moves themselves) with grab-and-move techniques (where the user moves the view) in VR, demonstrating that performance may depend on the user's spatial abilities. Although both methods they studied were active, we are particularly interested in the performance of passive navigation when combined with a predefined storyline.}

\section{Eliciting Design Considerations for Immersive Story Adaptations}
\label{sec:formative}

\begin{figure*}[t]
\centering
  \includegraphics[width=\linewidth]{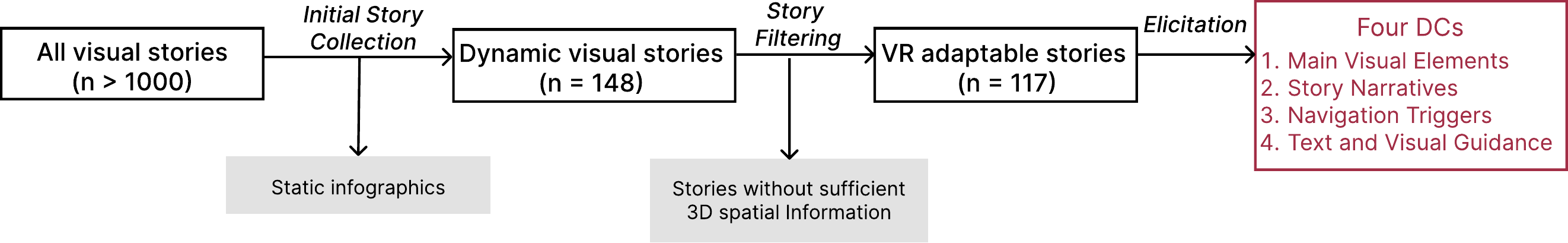}
  \caption{Formative Study Flow Diagram~\cite{page2021prisma}. We first collected dynamic visual stories from major news outlets. We then filtered out those without significant 3D spatial informations. With the remaining VR adaptable stories, we elicited four design considerations (DCs) for our later adaptations.}
  \Description{A process flow diagram showing the story filtering process for a VR study. The flow begins with over 1,000 visual stories, from which static infographics are filtered out. A total of 148 dynamic visual stories are selected through Initial Story Collection. These dynamic stories are further filtered, excluding stories lacking sufficient 3D spatial information, resulting in 117 VR adaptable stories. Finally, through Elicitation, four design considerations (DCs) are identified: Main Visual Elements, Story Narratives, Progression Triggers, and Text and Visual Guidance.}
  \label{fig:coding_procedure}
\end{figure*}

This project investigates the effects of viewpoint (ego vs. exo) and navigation (active vs. passive) in immersive storytelling through a user study. Preparing suitable study materials was the crucial first step.

{Visual storytelling can be characterized by four dimensions: narratives, transitions, structure, and the balance between explorability and explanability~\cite{stolper2018data, thudt2018exploration}. With various options in each category, visual storytelling can become quite complex. Since immersive storytelling is an emerging field with limited publicly available resources and few established design guidelines, we decided to develop our immersive stories.
Creating high-quality stories is a complex and challenging task. Therefore, instead of creating stories from scratch, we chose to adapt web-based stories from major news outlets. This approach allows us to ensure story quality while reducing potential confounding factors.
Considering the aforementioned four dimensions, we found that most stories from major news outlets follow a single-scene linear narrative structure with simple and continuous transitions and predominantly feature author-driven explanability. This simplicity is ideal for our initial exploration into immersive storytelling.
We further focused on stories with significant 3D spatial content due to their greater synergy with immersive environments. This approach enabled us to concentrate on the study itself, maintaining story quality while avoiding potential design biases.
}

To guide our adaptations, we collected existing 3D spatial stories from major news outlets and sought input from three VR experts. We compiled their insights into four generic design considerations, which were used for the adaptations discussed in \autoref{sec:adaptation}. An overview of the formative study process is shown in \autoref{fig:coding_procedure}.

In summary, the formative study and adaptation process, described in \autoref{sec:adaptation}, are essential steps for conducting the user study detailed in \autoref{sec:study}. \autoref{fig:procedure} illustrates the relationship between these steps.

The formative study involved three steps: 1) initial story collection, 2) filtering stories based on spatial information, and 3) eliciting design considerations for VR adaptation. Details of our collected stories are provided in the supplementary material.

\textbf{Initial story collection.}
Our first step is to collect abundant web-based stories. To ensure the quality of our visual story dataset, we collected stories from major news outlets and popular online media platforms, including The New York Times (NYT)~\cite{nyt}, The Washington Post~\cite{wapo}, Bloomberg~\cite{bloomberg}, The Guardian~\cite{guardian}, Reuters~\cite{reuters}, National Geographic~\cite{NG}, Financial Times~\cite{FT}, Los Angeles Times~\cite{lat} and The Pudding~\cite{pudding}. Some platforms feature dedicated visual story sections, which we prioritized, such as the Graphics section in NYT~\cite{nyt} and the Visual Stories section in The Washington Post~\cite{wapo}.

We focused on stories published between 2020 and 2024, as recent content tends to incorporate more interactive visuals. We excluded static stories without interactivity or visual transitions, as these did not align with the goals of our user study. 
{From the sources above, we browsed 1017 visual stories in total during the search process. Ultimately, our initial corpus comprised 148 visual stories that are either interactive or feature visual transitions. Most of them from The New YorK Times (102), followed by Reuters (16), The Washington Post (13), The Pudding (7), Financial Times (4), Los Angeles Times (2) and The Guardian (1).}

\textbf{Story filtering based on spatial information.}
Not all visual stories were suitable for our study. For example, animated bar charts may not offer a significantly different experience in an immersive environment compared to a desktop.

Since our goal was to investigate how the unique 3D capabilities of immersive environments impact the storytelling experience, we applied further filtering based on the presence of 3D spatial information. Stories were evaluated using two criteria: 1) sufficient 3D spatial content and 2) coherent spatial transitions within a unified environment. 
{The second requirement limited our scope to single-scene stories.}

For instance, a NYT story explaining COVID-19 transmission in a classroom was deemed suitable, as it contains rich 3D spatial information and smoothly transitions the viewpoint across different parts of the environment. In contrast, the NYT story ``Ukraine's Race to Hold the Line'' uses 2D visuals without enough 3D spatial content. Similarly, the story ``Seeing Earth from Outer Space'' includes 3D elements but presents multiple disjointed spaces, making each new scene appear completely separate from the previous one, complicating our intended study of immersive effects.

By applying these two criteria, we excluded 31 additional stories, resulting in a final set of 117 stories for our VR adaptation study.

\textbf{Elicitation of design considerations for VR adaptation.}
One of the key goals of our formative study was to derive design considerations that could inform the VR adaptation process. 
To accomplish this, we conducted an elicitation study with three experienced VR developers {whose backgrounds were medical VR, computer graphics in VR, and immersive animation design, respectively}. Each developer was assigned a roughly equal number of stories from our collection. They were asked to review their assigned stories individually and describe their intended adaptation in response to three specific questions related to our study factors, along with an open-ended question to explore additional design ideas. Developers were encouraged to provide detailed feedback.
\begin{itemize}
    \item Q1: What is the best way to view this story in an immersive environment? And why?
    \item Q2: Do you prefer to actively move in the 3D space or stand still and let the content move for you? And why?
    \item Q3: What other design considerations do you consider important?
\end{itemize}
We analyzed their responses using affinity mapping to identify common themes and patterns in their feedback.
\begin{figure*}[t]
\centering
  \includegraphics[width=\linewidth]{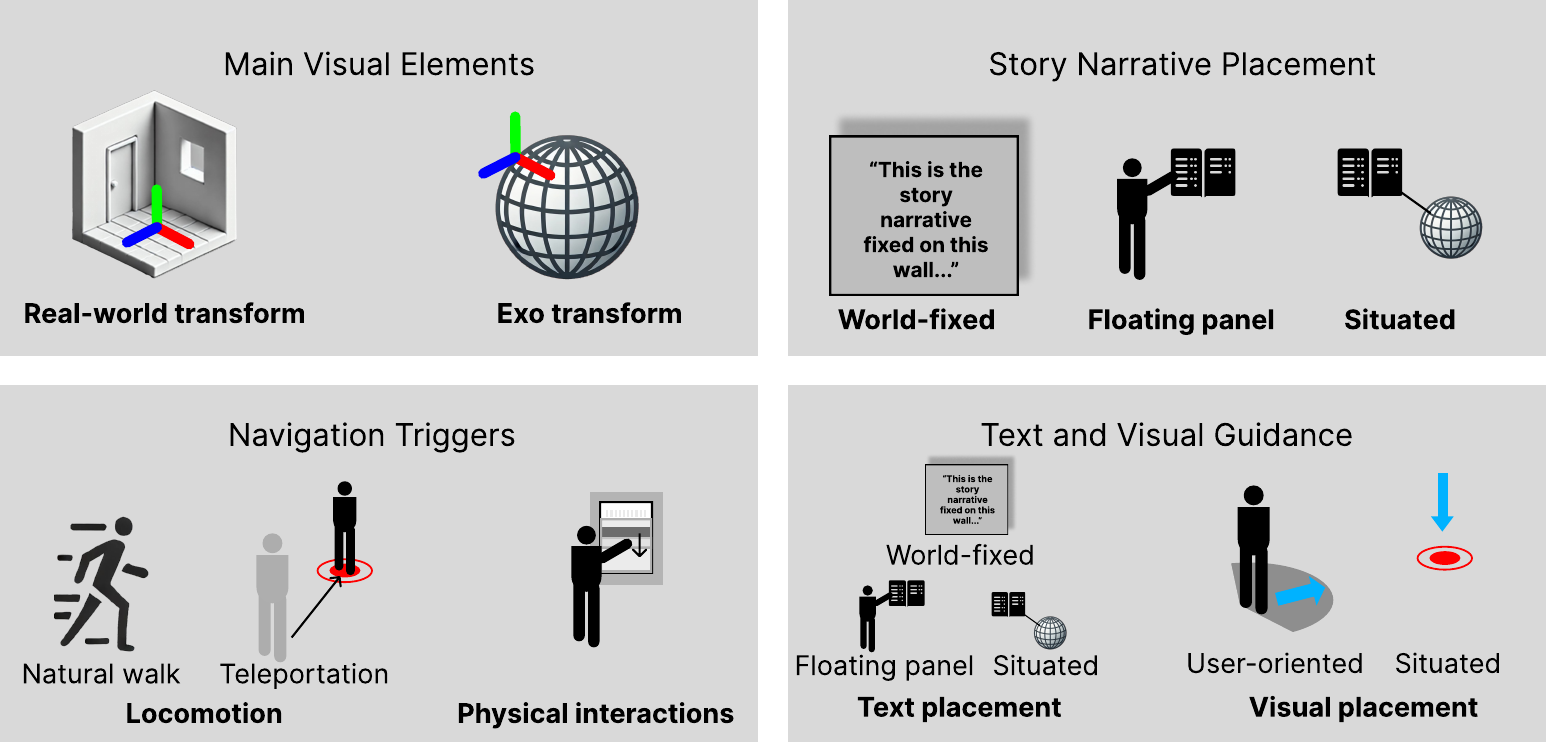}
  \caption{Four Major Story Components and Corresponding Available Options.}
  \Description{A diagram illustrating key design considerations for immersive storytelling in VR, presented in four categories: 1) Main Visual Elements: Real-world transform (elements positioned based on real-world coordinates) vs. Exo transform (global context positioning). 2) Story Narrative Placement: World-fixed (narrative anchored in the virtual world), Floating panel (text on a floating interface), and Situated (contextual narrative placement). 3) Navigation Triggers: Locomotion (natural walking or teleportation) and Physical interactions (interacting with objects like opening a window). 4) Text and Visual Guidance: Text placement (World-fixed, Floating panel, or Situated) and Visual placement (User-oriented vs. Situated for guiding attention).}
  \label{fig:design_options}
\end{figure*}
Regarding the viewpoint (Q1), \textit{information density} emerged as a key factor in choosing between egocentric and exocentric viewpoints. Developers generally favored an exocentric viewpoint for stories with dense spatial information—where a large amount of detail is concentrated within a limited area. For example, a story about the tall buildings in Manhattan, where the buildings are closely packed in a small space, would benefit from an exocentric perspective. Conversely, an egocentric viewpoint was preferred for stories featuring more dispersed spatial information, where the relative distances between objects are much greater than their individual sizes—such as a story about players on a football field. The developers largely agreed on the criteria for selecting viewpoints.

In contrast, there was no clear consensus on navigation design (Q2). For the same story, developers expressed differing preferences for active versus passive navigation. Some noted that passive navigation, which resembles the "scroll-to-progress" interaction commonly used on webpages, offers familiarity and ease of use. Others argued that active navigation in VR feels more intuitive and immersive, enhancing user engagement. These differing opinions were spread across various stories, without any strong patterns emerging.

In summary, while the developers reached a broad consensus on viewpoint selection, the decision between active and passive navigation remained inconclusive in our formative study.

Overall, we identified in the affinity diagram that experts frequently mentioned four major story components in their responses:
\begin{itemize}
    \item DC1: Main Visual Element. The main visual element serves as the centerpiece of the story, often represented by the primary 3D model or scene.
    \item DC2: Story Narratives. Story narratives refer to the structured sequence of textual content that guides the user through the story.
    \item DC3: Navigation Trigger. To trigger the navigation, the user needs to either move to a specific location or perform certain interactions.
    \item DC4: Text and Visual Guidance. Instructions regarding location and interaction information are necessary to guide the user in progressing through the story.
\end{itemize}

{Among these four design considerations, DC3 and DC4 hold unique importance for immersive stories. While web-based stories typically use straightforward navigation triggers, such as scrolling and clicking that require little to no guidance, immersive stories demand more explicit navigation and guidance cues due to the open-ended interactions audiences can perform. Therefore, it is essential to emphasize specific navigation triggers and provide corresponding guidance when adapting stories for immersive environments.}
These findings of our formative study provide structured guidelines for adapting web-based stories to immersive environments.

\section{VR Story Adaptation for User Study}
\label{sec:adaptation}

This section outlines the process of adapting well-established web-based stories into immersive environments for use as user study materials. We planned a within-subjects user study to minimize the influence of individual differences on the results. Given the four testing conditions in our study (as shown in \autoref{fig:procedure}), we required at least four immersive stories to mitigate the learning effect.

From the formative study results, we determined that viewpoint (ego vs. exo) was story-dependent, so we selected two stories for each viewpoint. 
Navigation (active vs. passive), however, was adaptable to any story, so we created both active and passive versions for each. A Latin square design was used to balance the presentation order of stories and navigation methods for participants. Four web-based stories were ultimately adapted into immersive formats for the study.

\subsection{Design Consideration Exploration}

Our goal was to explore design options for studying the effects of viewpoints and navigation. We examined the four design considerations identified in the formative study and selected the most appropriate options for adaptation.

For the {DC1} (\textit{main visual content}), we reused high-quality 3D models from the original stories. In egocentric stories, real-world scale was used for full immersion, while exocentric stories featured scaled-down models to fit within the user's field of view.

For the {DC2} (\textit{story narrative}), we reused the original text, considering several placement options:
1) Fixed placement: text appears in a consistent location (e.g., on a wall), though this may require frequent context-switching.
2) Floating panels: text is placed on movable panels, ensuring easy access with minimal effort.
3) Situated text: embedding text near the relevant objects to enhance semantic understanding.

{DC3} (\textit{navigation trigger}) in passive designs was triggered by simple actions like button presses. In active designs, it involved either locomotions~\cite{di2021locomotion} (e.g., walking or teleporting) or physical interactions (e.g., opening a window in VR).

For {DC4} (\textit{text and visual guidance}), concise textual instructions were provided, and visual indicators (e.g., arrows or hotspots) were used to guide navigation and interactions.

The design space of immersive stories is expansive, and building a comprehensive one is not the focus of this work and is beyond its scope. 
Our exploration of design options is not an exhaustive list of all possible alternatives but serves as structured guidelines for the necessary and critical adaptations. 
In the remainder of this section, we detail how we adapted the four selected stories using these design options.

\subsection{Case 1: How 911 changed Financial Distract in Manhattan}
\label{subsec:case1}
\begin{figure*}[t]
\centering
  \includegraphics[width=\linewidth]{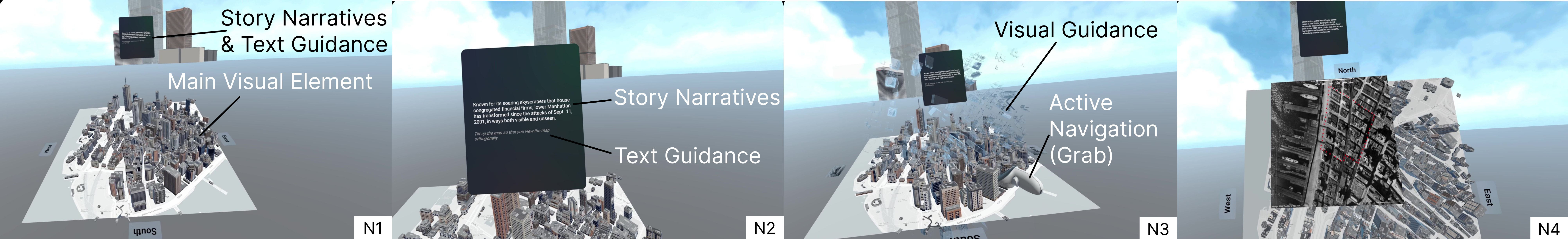}
  \caption{Illustration of Case 1. The main visual element is a Manhattan landscape with 3D buildings (N1). The story narratives and the text guidance for navigation are in a text panel above the main visual element (N2). A transparent copy of the main visual element periodically animates to a designated perspective, visualizing the text guidance (N3). Audiences follow the guidance and the story proceeds to the next paragraphs (N4)}
  \Description{A VR adaptation of a story about lower Manhattan's transformation post-9/11. It displays four screenshots from a VR environment where the story is adapted with interactive 3D elements. In the VR environment, participants engage with the following components: N1 (Main Visual Element, showing lower Manhattan in 3D), N2 (Story Narratives and Text Guidance displayed on a floating panel), N3 (Active Navigation, allowing users to manipulate the environment), and N4 (Historical imagery overlayed on the 3D map). The VR version integrates interactive spatial elements and narrative text for a more immersive experience.}
  \label{fig:911-VR}
\end{figure*}
This story covers the societal changes in Manhattan’s Financial District post-9/11~\cite{911}. The original story featured a 3D model of Manhattan’s densely packed buildings, making it ideal for an exocentric VR adaptation. {The story has 15 steps and 611 words in total. A step includes a paragraph of story narratives with corresponding visual elements. Audiences procceed to next step by following a specific navigation trigger.}

\textit{Main Visual Element.} We built a 3D landscape of Manhattan in VR (\autoref{fig:911-VR}-N1) through Mapbox~\cite{mapbox_2017_mapbox}. The landscape is 2m-by-2m square base map by default. The current building information came from mapbox.3d-buildings tileset. When the story narratives mentioned a particular group of buildings (e.g. WTC buildings), they are highlighted in blue.

\textit{Story Narratives Placement.} Story narratives are displayed on an interactive panel in VR (\autoref{fig:911-VR}-N2), which supports repositioning, resizing and scrolling. We utilized a floating panel design because the narratives are often associated with the whole main visual element. It is not situated to a specific parts.

\textit{Text and Visual Guidance.} Text guidance appeared in smaller gray italics to minimize distraction. Visual guidance, in the form of a transparent 3D model, periodically animated to show the next perspective audience should have to look at the landscape (\autoref{fig:911-VR}-N3).

\textit{Navigation Triggers.} Active navigation involved rotating and zooming in/out of the landscape to trigger changes in the scene. In passive navigation, the landscape automatically adjusted as participants progressed.

\subsection{Case 2: In the Atlantic Ocean, Subtle Shifts Hint at Dramatic Dangers}
\label{subsec:case2}
\begin{figure*}[t]
\centering
  \includegraphics[width=\linewidth]{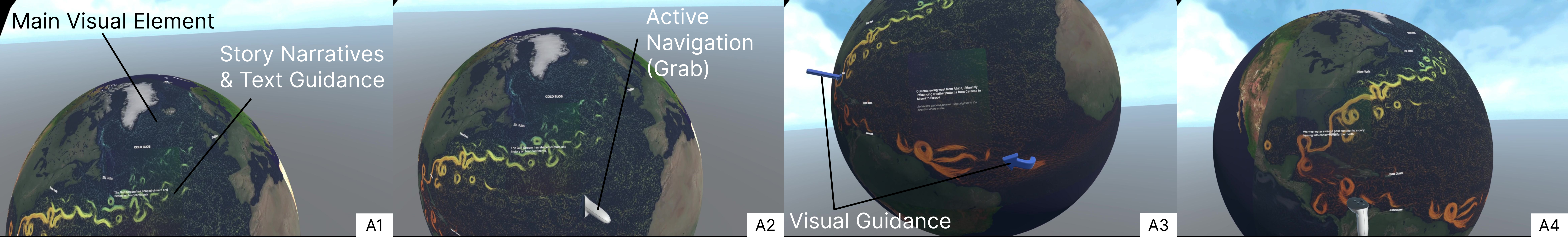}
  \caption{Illustration of Case 2. The main visual element of the story is a globe with animated current circulation in Atlantic Ocean (A1). The story panel with translucent background is fixed around the globe surface and centered between the audiences' view and the globe centroid (A1). Audiences actively navigate the story by grabbing the globe and rotating to the designated area mentioned in narratives (A2, A4), following the text and visual guidance (A1, A3).}
  \Description{A VR adaptation of a story about the weakening Gulf Stream in the Atlantic Ocean. It displays four VR screenshots: A1 (Main Visual Element, showing the Earth with a visual representation of the Gulf Stream and story narratives/text guidance), A2 (Active Navigation, where users can manipulate the globe for a better view), A3 (Visual Guidance, highlighting a specific region of the Gulf Stream), and A4 (Text and Visual Guidance, where users can interact with the 3D visualization to explore the story). The VR adaptation offers an interactive, spatial representation of the Gulf Stream data for a more immersive understanding.}
  \label{fig:atl-VR}
\end{figure*}
This story explores the impact of climate change on the Gulf Stream~\cite{atlantic}. We only selected the first half of the story, which has the most concentrated 3D spatial information. The main visualization of the original story was a globe with animated current circulations in Atlantic Ocean. Thus, we implemented a corresponding exocentric VR story. We summarized the text-only portion (484 words) into 2 paragraphs (100 words) to keep the length consistency. {The story has 15 steps and 503 words in total.}

\textit{Main Visual Element.} We adapted the original animation of ocean currents onto a 3D globe in VR. The globe’s size is adjustable to suit user preferences.

\textit{Story Narratives Placement.} The text portion of the story is in the same interactable panel as Case 1. However, in this case the panel is fixed around the globe surface and centered between audiences' view and the globe centroid (\autoref{fig:atl-VR}-A1). This is because the main visual element (i.e. the globe) occupies a much larger space than a normal human body and it would easily interfere with the panel if they were placed separately. To reduce the occlusion caused by the current placement, we make the panel background translucent ($\alpha = 0.1$, almost transparent) so that audiences can see underlying visualizations and keep track of the panel simultaneously.

\textit{Text and Visual Guidance.} The visual guidance of this case consists of 2 animated arrows (\autoref{fig:atl-VR}-A3). A bended arrow is placed in front of the audience's view on the surface and pointed to the next designated area mentioned in narratives. Another straight arrow is at that area and pointed to the globe centroid, indicating the direction audiences should look at the globe. It shows the next step 10 seconds after entering the current step. The text narrative is at the bottom of the panel and follows the same style as Case 1.

\textit{Navigation Triggers.} The active navigation of this case is primarily grabbing and rotating, as the original story rotated the globe along the Gulf Stream. For the 2 summarized text-only paragraphs, we adopt a step-away interaction: after audiences step away from the globe, the text panel leaves the surface and attaches to the audiences' view (0.5 meters from the eyes). Audiences can press a controller button to read the next paragraph and step closer to the globe after finishing 2 paragraphs. In the passive navigation, audiences only need to press a controller button and the globe would rotate and change the visualizations automatically.

\subsection{Case 3: What the Tulsa Race Massacre Destroyed}
\label{subsec:case3}
\begin{figure*}[t]
\centering
  \includegraphics[width=\linewidth]{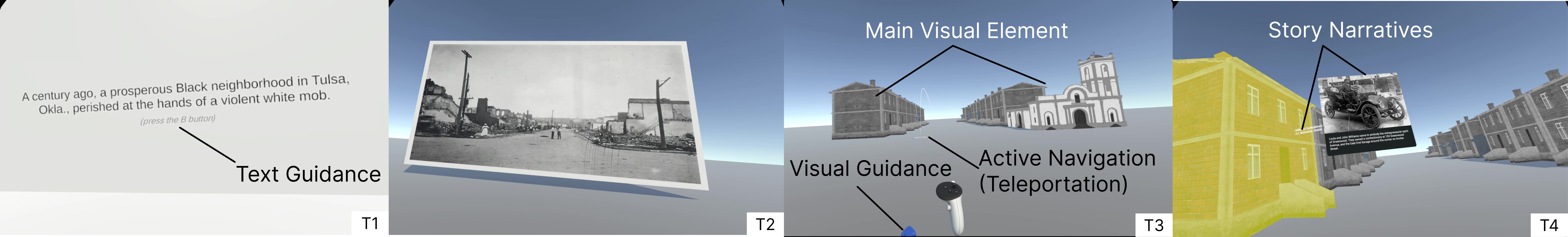}
  \caption{Illustration of Case 3. T1 and T2 replicated the signposting sentence and image at the beginning of the story. Audiences then actively navigate the story on a main avenue of Tulsa before the massacre, following the visual guidance of arrows and hotspots (T3). After they arrive at a building, the corresponding story narratives would appear in a panel and a label nearby (T4).}
  \Description{A VR adaptation of the story of the Tulsa Race Massacre. It features four VR screenshots: T1 (Text Guidance showing a narrative about the event), T2 (A historical photograph presented in the VR environment), T3 (Main Visual Element with active navigation using teleportation to explore the reconstructed neighborhood), and T4 (Story Narratives displayed as floating panels alongside the VR scene). The VR adaptation allows users to actively explore and interact with visual elements of the historical scene for an immersive learning experience.}
  \label{fig:tul-VR}
\end{figure*}
The story of this case illustrates the business achievement of black Americans in Tulsa, Oklahoma in the early 1900s and how the town was destroyed by racist mobs~\cite{tulsa}. We only selected the first half the story, which has the most concentrated 3D information. We summarized the second half of the story into a paragraph (130 words) and put it at the end so that story narratives are logically complete while keeping similar length with previous cases. The main visualization of the original story comprises a 3D reconstruction of Tulsa before the massacre, which leads to our design of a VR story with an egocentric viewpoint in a main avenue (Greenwood Ave) of Tulsa (\autoref{fig:tul-VR}-T3). {The story has 18 steps and 618 words in total.}

\textit{Main Visual Elements.} Greenwood Ave of Tulsa included a number of black businesses alongside the road. Audiences follow a predefined route, in which one building would be highlighted at each step. At the beginning of the story, we keep the signposting sentence and image in the original story (\autoref{fig:tul-VR}-T1, T2) as they concisely describe the main event of the story.

\textit{Story Narratives Placement.} The narratives of this case consist of two parts: business labels and introductory paragraphs. In the original story, they were presented in two passes of the avenue. In our VR story, we combine these two types of narratives into one pass because it is more natural to follow a linear order, which ends the story when audiences reached the end of the avenue. Each building has a label listing the businesses inside, which is shown close to the building. The introductions are not on every building as not every business was introduced in the original story. They would be shown in front of the audiences' views.

\textit{Text and Visual Guidance.} The text guidance of this case only exists at the beginning (\autoref{fig:tul-VR}-T1), which asks audiences press a controller button to proceed. After audiences enter the Greenwood Ave, only the visual guidance indicates the building audiences should go. It shows the next step 10 seconds after entering the current step. It includes an arrow fixed at the bottom of the field of view (FOV), a hotspot highlight of the destination and another arrow above the hotspot pointing downwards. We fix an arrow in the FOV because it ensures audiences to find the hotspot even if it is behind them.

\textit{Navigation Triggers.} The active navigation of this case is predonimantly teleportation. Audiences follow the visual guidance and teleport to the front of each building and gradually to the end of the avenue. In the passive navigation, audiences only need to press a controller button and they are translated to the destination automatically.

\subsection{Case 4: Why Opening Windows Is a Key to Reopening Schools}
\label{subsec:case4}
\begin{figure*}[t]
\centering
  \includegraphics[width=\linewidth]{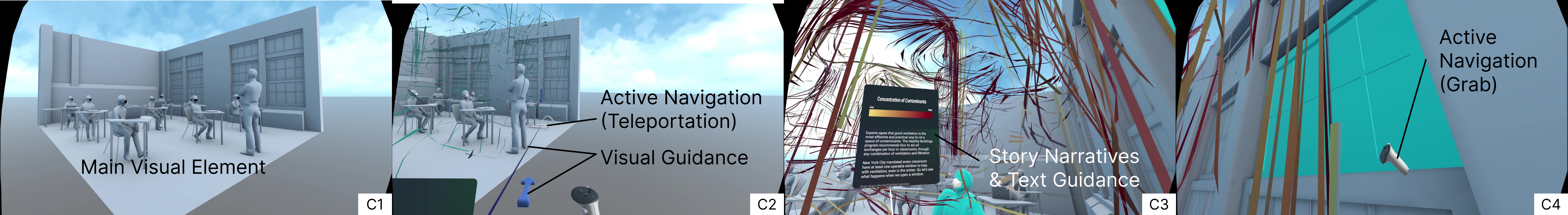}
  \caption{Illustration of Case 4. The main visual element was a 3D classroom model (C1). Audiences actively navigated to the next designated area mentioned in narratives by teleportation (C2). After they arrived, they saw a visualization of airflow with corresponding narratives on the panel (C3). They were asked to drag down the window to see a visualization of a different airflow (C4).}
  \Description{A VR adaptation of a story on the importance of opening windows for safe school reopening. It features four VR screenshots: C1 (Main Visual Element showing the virtual classroom with a seated arrangement of students), C2 (Active Navigation via teleportation with visual guidance lines showing airflow), C3 (Story Narratives and Text Guidance displayed alongside airflow visualizations), and C4 (Active Navigation through object manipulation, with users grabbing and interacting with windows to alter the airflow in the classroom). The VR version allows for an interactive, spatial understanding of how air circulates within the environment and the impact of opening windows.}
  \label{fig:cov-VR}
\end{figure*}

The story of this case illustrated the importance of ventilation in classroom during COVID pandemic by comparing the concentrations of contamination in different ventilation conditions. The original story visualized the concentrations in both airflow and heatmap. We only selected the airflow visualization as it contains more 3D spatial information than the heatmap. Although the original story adopted a exocentric viewpoint, we decided to implement an egocentric VR story because it would be more immersive to be in a life-sized room rather than viewing a smaller room model. {The story has 13 steps and 513 words in total.}

\textit{Main Visual Element.} The classroom includes some models of students and a teacher. An air circulation animation is shown to depict the inhale and exhale flow in the classroom. A student is then highlighted as infected (in cyan in \autoref{fig:cov-VR}-C3), and the airflow becomes yellow and red to indicate the concentration differences. The interactable objects (e.g. the cyan window in \autoref{fig:cov-VR}-C4) are also highlighted when interacting with them to proceed through the story.

\textit{Story Narratives Placement.} The story narratives are mainly in the interactive panel. In this case, the panel is attached to the left controller, as a large portion of the narratives are about the airflow, which occupy the whole classroom and do not have a specific area to attach to.

\textit{Text and Visual Guidance.} Following Case 1 and 2, the text guidance is also at the bottom of the panel with the same font. The visual element follows the same design as Case 3, and we add an extra line connecting the hotspot and the bottom of the panel because the model of students and teacher might occlude the hotspot on the ground. Similiarly, it shows the next step 10 seconds after entering the current step.

\textit{Navigation Triggers.} The active navigation of this case involves teleportation and grabbing. Audiences are asked to teleport to different locations of the classroom to view the airflows. They can also drag down the window and put on the fan to change the ventilation conditions. In the passive navigation, audiences only need to press a controller button and their viewpoint, ventilation condition and corresponding visualizations would change automatically.

{\subsection{Story Caliberations}}
\label{sec:calibration}
{The original four stories have different story length, visual complexity and spatial information, which is hard to compare directly. We calibrated four cases based on the first three design considerations (DC1-3 in \autoref{sec:formative}): main visual element, story narratives, and navigation triggers. For text and visual guidance (DC4), a ``one-size-fits-all'' design could negatively impact some story instances. Therefore, we opted to use designs optimized for each individual story instance.}

{DC1: \textit{Main Visual Element(s)}. We calibrated the size of main visual element between two exo cases (Case 1 and 2) and two ego cases (Case 3 and 4), respectively. For exo cases, the main visual element was bounded by a cube with 2m edges. For egocentric cases, the main visual elements spread in a 20m-by-20m square arena, and the size of each elements was adjusted accordingly. Due to the different stories of each case, we were not able to have similar number of visual elements between exo and ego cases. But we guaranteed that the number of visual elements audiences needed to pay attention to at each step was no larger than 5.}

{DC2: \textit{Story Narratives}. For each case, we divided the story into about 15 steps. The steps were created based on the original story narrative setups, which consisted of several short paragraphs in a scrollytelling format. We combined neighboring short paragraphs and summarized long paragrahs to make each step about 30 to 50 words long. Thus, four cases had the similar lengths of narratives (15$\pm$3 steps and 550$\pm$50 words).}

{DC3: \textit{Navigation Triggers}. The navigation triggers in two exo cases were looking at the main visual element from a specific angle. In two ego cases, they were teleporting to a specific location. Due to the different scene and visual element settings, we cannot use the same navigation triggers between exo and ego cases.}

{DC4: \textit{Text and Visual Guidance}. The effectiveness of text and visual guidance was closely related to the other 3 DCs, particularly the main visual elements. Thus, we tried to optimize the guidance by designing the guidance separately based on visual element setup of each story case. For example, to indicate the manipulations audience needed to do (primarily rotation), Case 1 used the animations of a translucent "ghost" map and Case 2 used two arrows. Animation provided the clearest instructions of the next step, but it was not applicable to Case 2 whose main visual element was a globe. Similarly, both Case 3 and 4 used arrows and an on-ground hotspot to indicate the position audiences needed to teleport to. However, the hotspot could be easily occluded by student and desk models in Case 4. So we used an extra line connecting the hotspot to the left controller so that audiences would easily find the next position to teleport. We acknowledged that such inconsistency might cause confounding factors, but no single text and visual guidance design could serve as a clear and effective road sign in all four cases.}

\section{User Study}
\label{sec:study}

With the four adapted VR stories, we conducted a user study to evaluate the performance of spatial immersion and understanding among different settings of viewpoints and navigations.
Our study was approved by our institution's IRB. 
It implements a within-subject comparison where each participant experiences active and passive navigation methods in both egocentric and exocentric viewpoints. 
Each session lasted about 2 hours and participants were compensated with \$20 for their time.

\subsection{Study Factors and Conditions}

\begin{figure*}[t]
\centering
\begin{subfigure}{0.45\linewidth}
  \includegraphics[width=\linewidth]{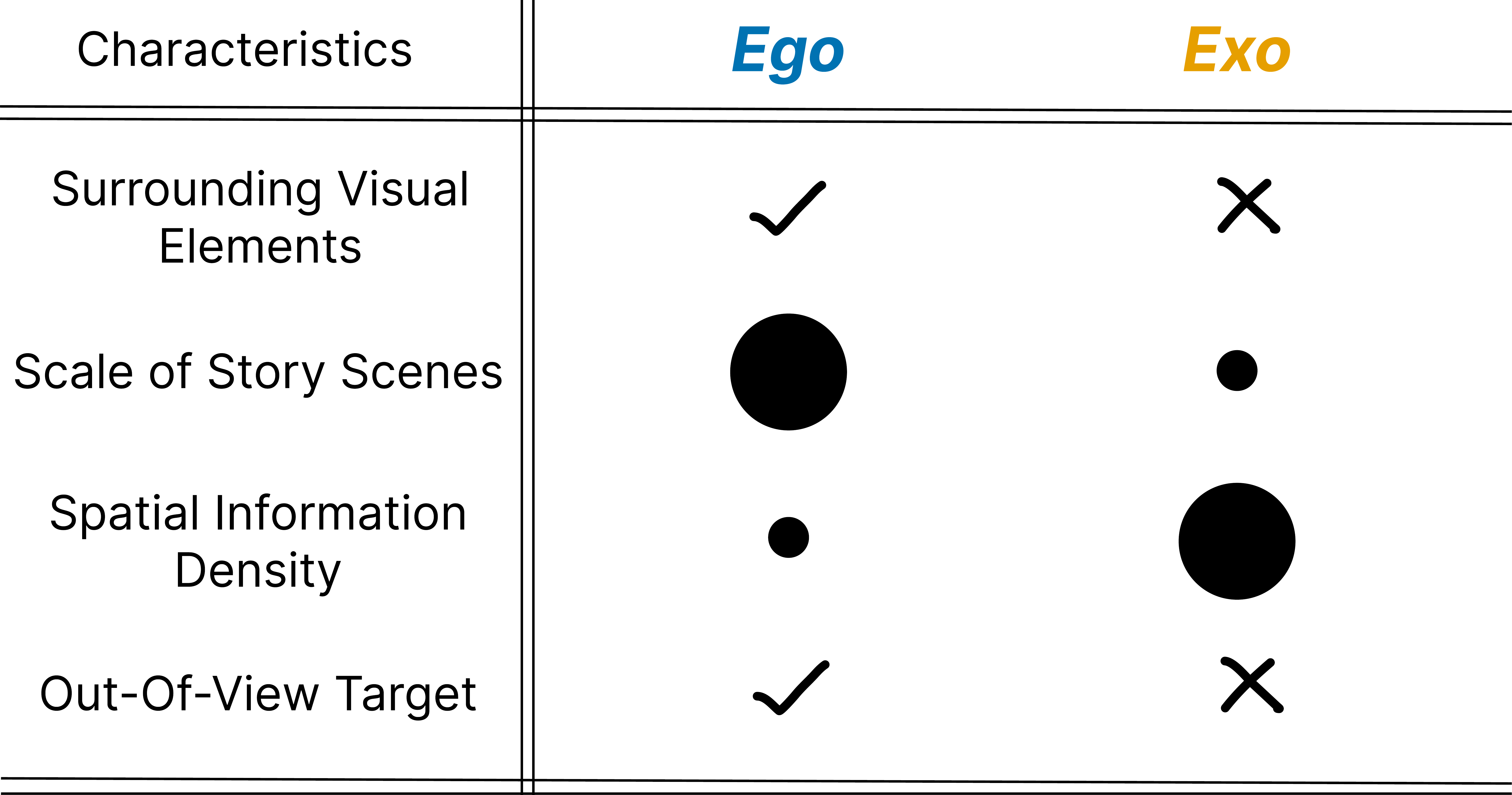}
  \caption{}
  \label{fig:view}
\end{subfigure}
\begin{subfigure}{0.45\linewidth}
  \includegraphics[width=\linewidth]{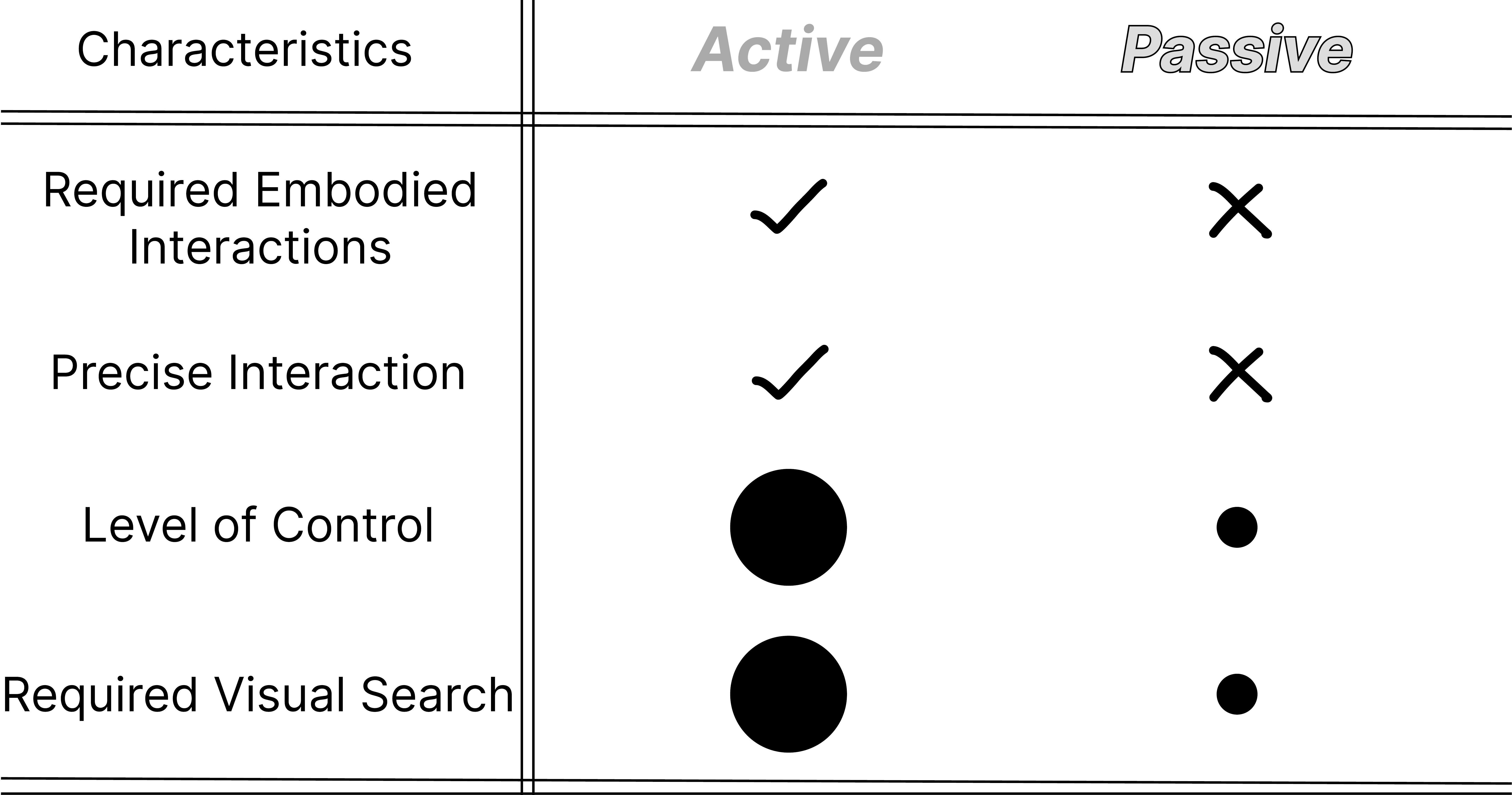}
  \caption{}
  \label{fig:nav}
\end{subfigure}
  \caption{Characteristics of Four Conditions. A tick mark indicates that the condition has corresponding characteristics, and the cross mark indicates the absence. If both conditions have the characteristics, they are labeled with circles with size indicating the extent (more or less).}
  \label{fig:characteristic}
  \Description{A comparison table outlining the characteristics of egocentric (Ego) vs. exocentric (Exo) viewpoints and active vs. passive navigation in immersive storytelling. On the left, the table compares Ego and Exo based on four characteristics: surrounding visual elements (Ego includes surrounding elements while Exo does not), scale of story scenes (Ego has larger scenes), spatial information density (higher in Exo), and out-of-view target visibility (available in Ego but not Exo). On the right, the table compares Active and Passive navigation in terms of required embodied interactions (more in Active), precise interaction (more in Active), level of control (higher in Active), and required visual search (more in Active than Passive). Larger circles indicate greater intensity or significance of the characteristic.}
\end{figure*}

We studied the Ego and Exo viewpoints as well as active and passive navigations, including their individual and combined effects. \autoref{fig:characteristic} shows the characteristics of the four conditions.

\textit{Ego vs. Exo Viewpoint.} The differences between Ego and Exo lies primarily in the visual representations. For example, Ego stories have surrounding visual element around audiences, while visual elements in Exo stories are primarily in front of the participants. There are also different scales of story scene, as Ego stories tend to have a much larger range compared to the space a normal person can occupy. Additionally, the information density in audiences' view is low, but there are out-of-view objects that they may leave out. Exo stories, in contrast, have everything in audiences' view, which increases the information density.

\textit{Active vs. Passive Navigation.} The differences between active and passive are more distinct. Active stories require embodied interactions to push forward the story, and oftentimes the interactions requires precision (e.g. going to a specific location or looking from a specific angle. In passive stories, only a minimum effort (i.e. clicking a button on the controller) is required. Moreover, active stories give audiences a higher level of control by asking them to proceed with story manually. This also increases the amount of visual search necessary for identifying the next step. In passive stories, visual elements transition to their next predefined state automatically, and audiences have less control and visual search in this process.

To systematically test the main factors (i.e., viewpoint and navigation), we include four conditions in our study:
Ego+Active, Ego+Passive, Exo+Active, and Exo+Passive (illustrated in \autoref{fig:teaser}).

\subsection{Study Setup}
Our user study was conducted in our lab space at the university.
Participants experienced our stories using a Meta Quest 3 virtual reality headset, which has a pixel resolution of 2064 x 2208 per eye and 90Hz refresh rate.
The headset was connected to PC via Air Link, allowing it to wirelessly leverage PC computing power.
Participants were allowed to physically move and interact with stories in a 3x3$m^2$ area.
During the study, we set the controllers as the only input devices for stability. We mapped the grip press to the grab interaction and thumbstick forward push to teleportation for both controllers.
This study setup was introduced at the very beginning of the study session for each participant.

\subsection{Participants}
We recruited participants from our university by sending out recruitment emails to mailing lists and Slack channels and selected 24 participants (11 females, 13 males, aged 21 to 31) from all responses.
11 participants were experienced in XR, and 5 were experienced in visualization/storytelling. 3 participants had both expertise.
Participants came from various backgrounds, including computer science, UX design, aerospace engineering, and Human-Computer Interaction.

{As ~\autoref{sec:calibration} mentioned, all four cases were calibrated in story length and overall design languages. 
Each participant experienced the four cases, with the order counterbalanced using a Latin square matrix (ego vs. exo)$\times$(active vs. passive).
Since each story has both an active and a passive version, there were a total of eight possible sequences. Further details can be found in our supplementary materials.
We acknowledged that this within-subjects setting could cause possible learning effects or fatigue, but we believe the difference in the nature of each cases minimizes these confounding factors.}

\subsection{Study Procedure}
Our study procedure includes four major steps:

\noindent{}\textbf{1. Introduction:} We first showed participants the consent form for the study. After they read and signed the form, we introduced the concept of immersive storytelling, ego/exocentric viewpoints, and active/passive navigation methods.

\noindent{}\textbf{2. Training:} We built a training scene that helps participants learn necessary interactions (e.g., grab, scale, teleportation) and story setup (e.g., narrative text panels, visual guidance design) in our VR stories. 
The training scene is a mockup of four real stories that use the same design language.

\noindent{}\textbf{3. Experiencing the story:} After the training session, participants were to experience the four VR stories. None of them had read the original story before. After participants finished each story, we asked them to complete 3 tasks: (a) completing a survey of presence and perceived workload of experiencing the VR story; (b) describing their understanding of the story's main idea and important details; and (c) drawing the story scene. The data we want to collect through these tasks will be described in Sec \ref{subsec:DataCollection}.

\noindent{}\textbf{4. Final interview:} Finally, after participants experienced all the stories, they were asked to discuss the differences among the four stories in terms of their engagement, help of understanding and other user experiences. 
The purpose of this interview was to collect subjective feedback on each story and find out their common considerations.

\subsection{Task Data Collection}
\label{subsec:DataCollection}
Through the study, we aimed to investigate the impact of ego/exocentric viewpoints and active/passive navigation on spatial immersion and understanding of VR stories. 
For spatial immersion, we collected responses to the Presence Questionnaire \cite{berkman2021group} and NASA-TLX \cite{hart1988development} to measure spatial presence and perceived workload, respectively. 

To evaluate the understanding, we focused on the story content comprehension and spatial information depiction. 
The total and categorical points of these two tasks for each story was the same since the their story lengths are similar.
For each story. we designed a set of comprehension rubrics that incorporated three sets: (a) \textit{main idea memorization} (b) \textit{detail memorization} (e.g. numbers, names, quotes or definitions, etc.); and (c) \textit{random inspirations} (e.g. personal opinions, analyses or conclusions that are not presented directly in the story content).
The rubrics is attached in the supplementary materials.

We asked participants to describe their understanding of the main idea and important details, which was then compared to our rubrics. 
We counted the number of facts in the rubrics that were mentioned by participants as their scores.
We also recorded the facts that do not appear in the rubrics.
If a fact was frequently mentioned (e.g., more than half of the participants mentioned it), we reconsidered the importance of this fact and added it to the rubrics if the majority of the study team thought it was important.

As for the spatial information depiction task, we asked participants to draw a picture of the story scene to measure their memorability of spatial details. We specifically mentioned 2 requirements: (a) \textit{object existence}: participants should capture as many spatial elements in the story as possible, including the objects, annotations, and the viewer path (the change of attention); and (b) \textit{object transform}: for each spatial element, provide its position, rotation/orientation and scale/size as accurate as possible.
The rubrics is attached in the supplementary materials.

We did not expect participants to draw the exact shape of elements.
They were encouraged to use basic geometries (e.g. lines and rectangles) and text descriptions to depict the elements.

Finally, we logged the story session duration. This was to understand if the four combinations of conditions would cause any differences in the time audiences stayed in the immersive stories.

\subsection{Hypotheses}
\label{subsec:hypotheses}
Based on the characteristics in \autoref{fig:characteristic}, we proposed the following hypotheses.

\textbf{Story Session Duration.}
We hypothesize that active navigation will result in longer story session durations compared to passive navigation. The interactive elements in active navigation require users to engage physically with the environment, adding time to the experience. Prior studies have shown that increased interaction in VR generally extends session time due to higher engagement levels and the need for precise actions in immersive environments~\cite{lages2018move, servotte2020vr}. This additional engagement likely leads participants to spend more time exploring the environment and interacting with the content, especially in active conditions.

Additionally, Ego viewpoints are likely to further increase session duration, as spatial navigation in egocentric environments tends to be more complex and challenging, requiring users to explore the scene more actively, which can lead to extended engagement~\cite{van2018virtual}.

\textbf{Story Understanding.} 
We hypothesize that active navigation will outperform passive navigation on both content comprehension and spatial information depiction, as embodied interaction has been shown to improve cognitive engagement and memory retention~\cite{wilson2002six, makransky2019adding}. Active navigation encourages physical movement, which enhances attention to both content and spatial information, leading to better comprehension and recall~\cite{servotte2020vr}. 

We also expect that exocentric viewpoints will result in better content comprehension because they provide an overview of the entire scene, facilitating the integration of key story elements and improving factual recall~\cite{yang2018maps}.
Egocentric viewpoints will lead to better spatial understanding, as being immersed in the scene helps participants build stronger spatial awareness and memory, which is supported by the "memory palace" technique~\cite{legge2012building, krokos2019virtual}.

\textbf{Spatial Immersion.} 
We expect that Ego viewpoints combined with active navigation will result in the highest levels of spatial immersion. Active navigation increases the sense of presence and control, while Ego viewpoints create a fully immersive experience by surrounding participants with 3D elements~\cite{slater2009place}. Prior studies suggest that immersion is enhanced when users feel directly involved in the environment, with Ego viewpoints offering a more embodied experience~\cite{cummings2016immersive}.

While active navigation may increase perceived workload due to the effort required to interact with the environment, we hypothesize that the enhanced sense of presence and engagement outweighs the additional cognitive demand, especially when paired with the Ego viewpoint~\cite{hart1988development}. This combination offers an optimal balance between immersion and interactivity, despite a potential trade-off with workload~\cite{servotte2020vr}.
\section{Results}
In this section, we reported the study results.
We performed the statistical analysis to test significances on the comprehension and rating data collected from three after-session tasks and subjective comments from final interviews. Significance values were reported for $p < 0.1(.)$ (dashed line in figures), $p < 0.05(*)$, $p < 0.01(**)$, and $p < 0.001(***)$.

\subsection{Story Session Durations}
\begin{figure}[h]
\centering
  \includegraphics[width=\columnwidth]{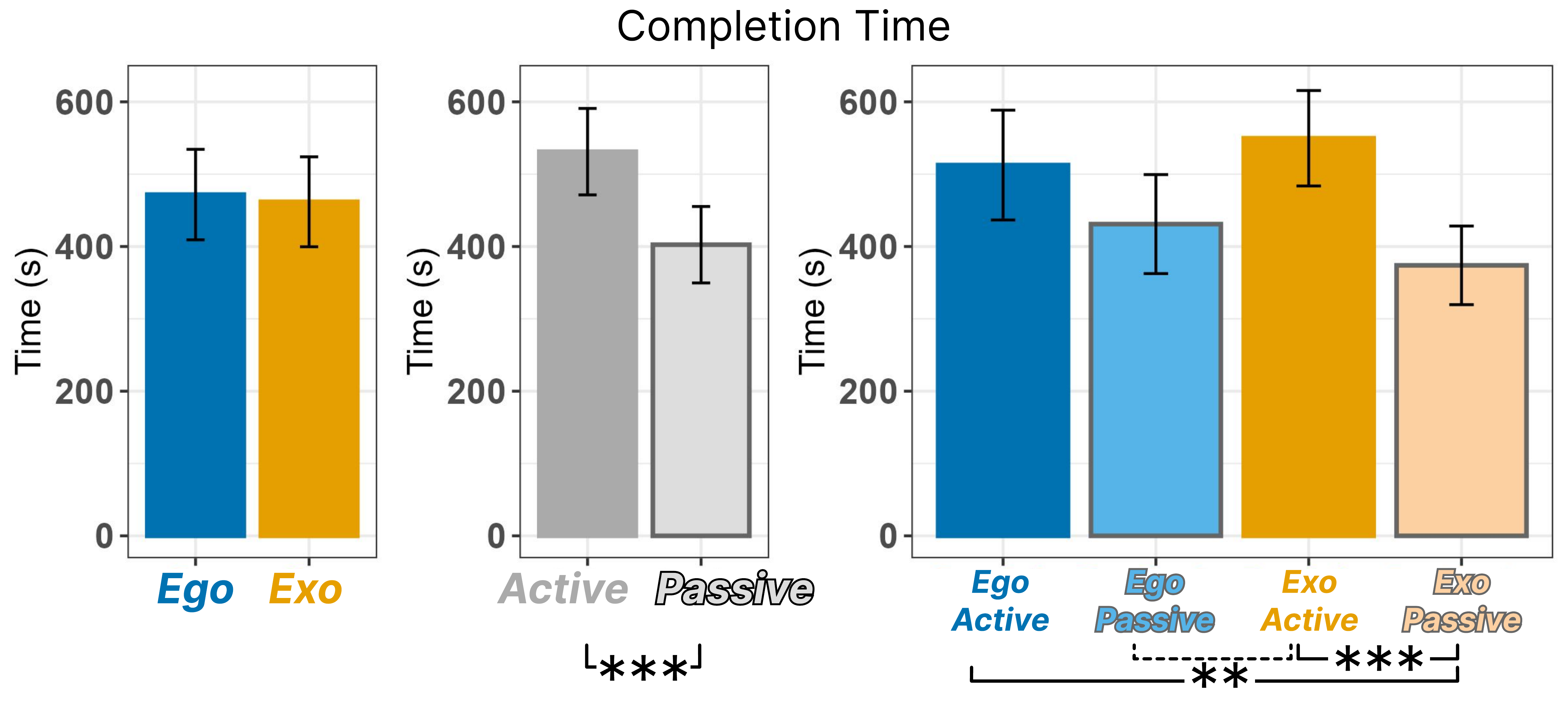}
  \caption{Average Story Session Durations. The dashed line indicates the marginal significance ($p < 0.1$) and the solid line indicates statistical significance ($p < 0.05$).}
  \label{fig:res_time}
  \Description{A bar chart comparing the completion time (in seconds) for different combinations of viewpoint (Egocentric vs. Exocentric) and navigation style (Active vs. Passive) in a VR storytelling study. The first chart shows similar completion times for Ego (blue) and Exo (orange) viewpoints, averaging around 450 seconds. The second chart shows a significant difference between Active (dark gray) and Passive (light gray) navigation, with Active taking longer (~550 seconds) compared to Passive (~400 seconds), as indicated by three asterisks *** denoting high statistical significance. The third chart breaks down completion times by both viewpoint and navigation combinations, with Ego Active taking the longest (~550 seconds), followed by Exo Active, while Passive conditions (Ego Passive and Exo Passive) take less time (~400 seconds). Significant differences between conditions are marked with two ** or three asterisks *** indicating different levels of statistical significance.}
\end{figure}

We collected the duration of each story session, defined as the total time the headset was worn, excluding any pauses. It is shown in \autoref{fig:res_time}. 
{\textbf{Participants spent significantly more time on active navigation cases than passive ones, while the duration between ego and exo viewpoints was not salient.}} 
A Shapiro-Wilk normality test confirmed the data followed a normal distribution ($p < 0.05$). We ran a General Linear Mixed Model (GLMM) with a Gaussian link function, using \textit{Viewpoint} (ego/exo), \textit{Navigation} (active/passive), their interaction ($\textit{Viewpoint} \times \textit{Navigation}$), and participant \textit{Group} as fixed effects, with participant ID as a random effect. Results showed significant effects only for \textit{Viewpoint} ($p < 0.001$), and a pairwise post-hoc Tukey's Honest Significant Difference (HSD) test confirmed a significant difference in duration between active and passive conditions in the exo viewpoint.

\subsection{Story Understandings}
We assessed story understanding through content comprehension and spatial information depiction tasks. 
{We found that \textbf{ego viewpoint showed significantly better performance on spatial information depiction.}}

\textit{Content Comprehension.}
\begin{figure*}[h]
\centering
  \includegraphics[width=\linewidth]{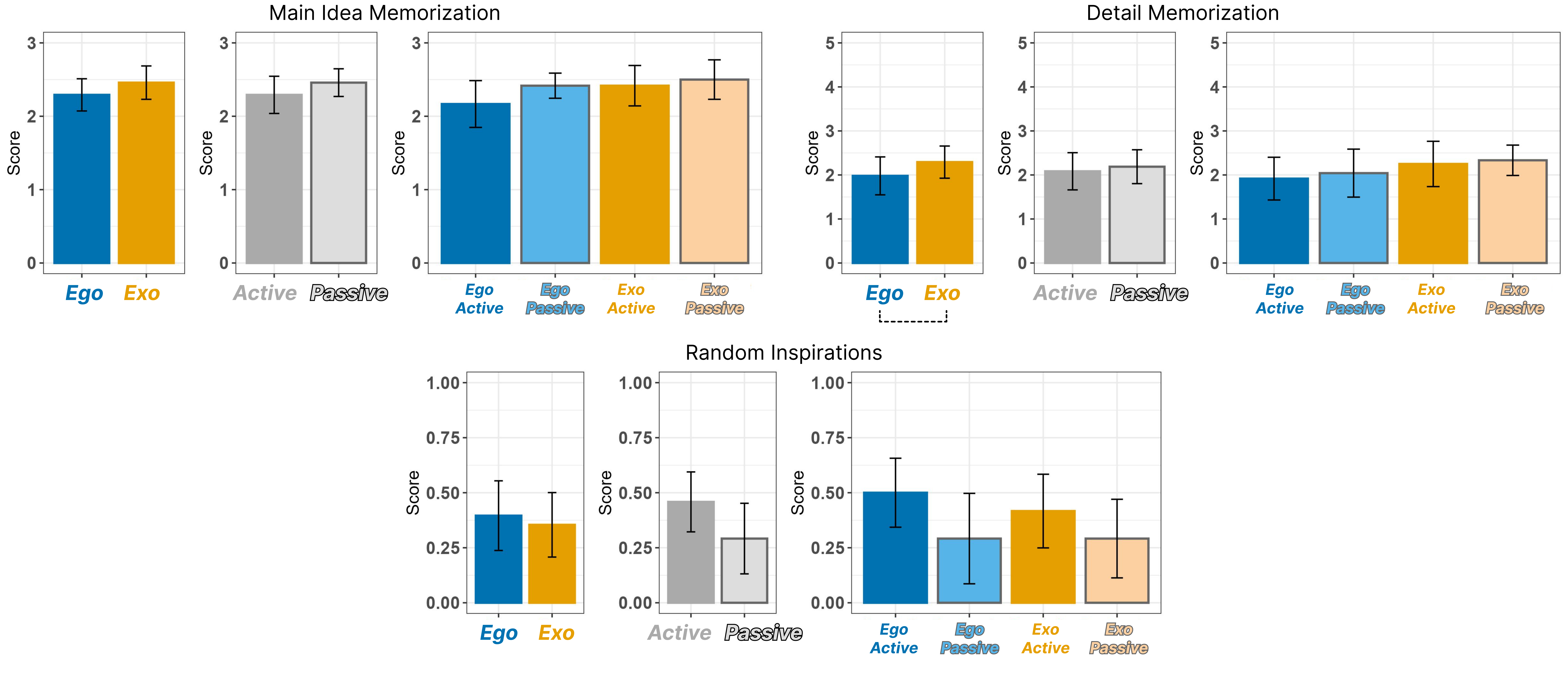}
  \caption{Average Sub-Scores of Content Comprehensions. The dashed line indicates the marginal significance ($p < 0.1$).}
  \label{fig:res_comprehension}
  \Description{A series of bar charts displaying scores for Main Idea Memorization, Detail Memorization, and Random Inspirations in relation to egocentric (Ego) vs. exocentric (Exo) viewpoints and active vs. passive navigation in a VR storytelling study. For Main Idea Memorization, the first row of charts shows no significant difference between Ego and Exo, or Active and Passive, with both averaging similar scores. When broken down by viewpoint and navigation, Exo Passive has the highest score, while Ego Active and Exo Active also perform well. For Detail Memorization, there is no marked difference between Ego and Exo or Active and Passive, with all combinations performing similarly. For Random Inspirations, the charts show slightly higher scores for Ego over Exo, and Active over Passive. Ego Active has the highest score, suggesting this combination encourages more creative inspiration. Error bars indicate standard deviation across the different conditions.}
\end{figure*}
The content comprehension task consists of three sets of questions: main idea memorization, detail memorization, and random inspirations. They have a total score of 3, 5 and 1, respectively. Each score corresponds to an idea or fact in the first 2 sets. For the random inspiration, participants would get the score if they mentioned anything derived from the story (e.g., linking to their own experiences or adding their own opinions). The only case in which they would lose this point is when they simply restated the story content (e.g. ``I learned [idea/fact of the story] is important'').

As \autoref{fig:res_comprehension} shows, the exo and passive conditions had a higher memorization performance, while the active condition achieved better random inspiration. In terms of the four combinations, exo$+$passive was the best on two memorizations, whereas ego$+$active had the highest score in random inspirations.
We then tested the normality of the data using a Shapiro-Wilk normality test, which showed that the data followed a normal distribution ($p < 0.05$).
Thus, we ran a GLMM with a Gaussian link function.
The results showed only a marginal significance of \textit{Viewpoint} on detail memorization ($p = 0.09$).

\textit{Spatial Information Depiction.}
\begin{figure*}[h]
\centering
  \includegraphics[width=\linewidth]{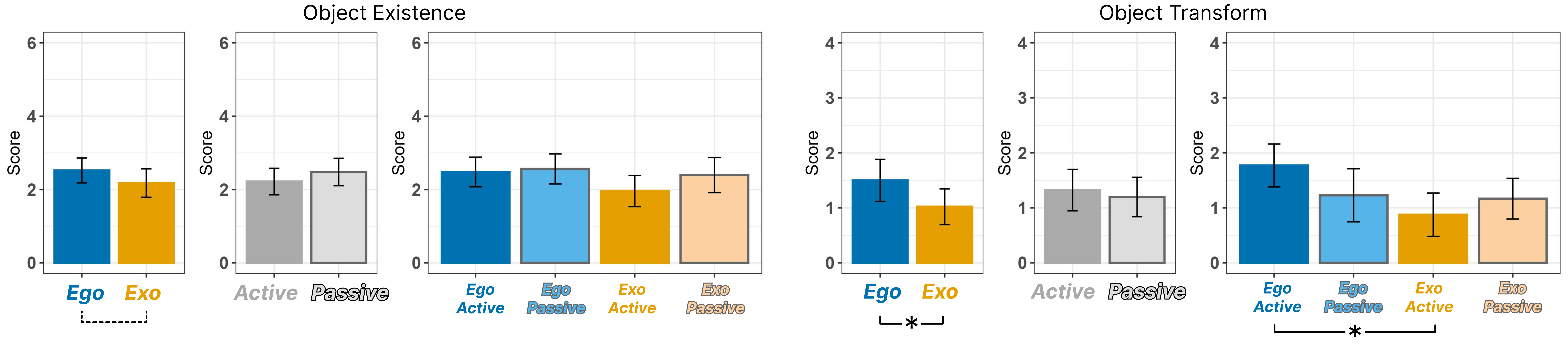}
  \caption{Average Sub-Scores of Spatial Information Depiction. The dashed line indicates the marginal significance ($p < 0.1$) and the solid line indicates statistical significance ($p < 0.05$).}
  \label{fig:res_drawing}
  \Description{A series of bar charts showing scores for Object Existence and Object Transform tasks in a VR storytelling study comparing egocentric (Ego) vs. exocentric (Exo) viewpoints and active vs. passive navigation. For Object Existence, the first set of charts shows no significant difference between Ego and Exo or Active and Passive, with all conditions scoring similarly around 2-3 points. In the breakdown by viewpoint and navigation, Exo Passive has the highest score, but differences remain minimal. For Object Transform, the second set of charts reveals a significant difference between Ego and Exo, with Ego scoring higher, particularly in Ego Active, which has the highest score. A single asterisk * indicates statistical significance in the difference between these conditions. Error bars represent standard deviations for each condition.}
\end{figure*}
We evaluated participants' drawings for object existence and object transform.
\autoref{fig:res_drawing} showed ego outperformed exo, especially in the ego+active condition.
We also observed that participants had the worst memorability of spatial information in exo$+$active.
We then tested the normality of the data using Shapiro-Wilk normality test, which showed that the data followed a normal distribution ($p < 0.05$).
Thus, we ran a GLMM with a Gaussian link function.
Results showed that only \textit{Viewpoint} has marginal significance on object existence ($p = 0.09$) and significance on object transform ($p < 0.05$). It also showed the signifcance of $\textit{Viewpoint}\times\textit{Navigation}$ on object transform ($p < 0.05$).
A post-hoc Tukey's HSD test of $\textit{Viewpoint}\times\textit{Navigation}$ on object transfrom showed that ego$+$active has a significantly higher score than exo$+$active.

{\textit{Result Summary.} Our hypothesis was partially supported by the results. We observed significantly better spatial information depiction in cases using the ego viewpoint. However, the results did not support the hypothesis that the exo viewpoint leads to better content comprehension, and navigation did not play a significant role in story understanding.}

\subsection{Spatial Immersion}
We collected the subjective ratings on spatial immersion for each case using a Presence Questionnaire (PQ) and NASA-TLX. We found that \textbf{active navigation enhances the sense of presence compared to passive navigation. It does not greatly increase workload when combined with ego viewpoint.}

\textit{Sense of Presence}
\begin{figure*}[t]
\centering
  \includegraphics[width=\linewidth]{Figures/presence.pdf}
  \caption{Average Sub-Scores of Presence Questionaire. The dashed line indicates the marginal significance ($p < 0.1$) and the solid line indicates statistical significance ($p < 0.05$).}
  \label{fig:res_presence}
  \Description{A series of bar charts illustrating scores for Realism, Possibility to Act, Quality of Interface, Possibility to Examine, and Self-Evaluation of Performance in a VR storytelling study, comparing egocentric (Ego) vs. exocentric (Exo) viewpoints and active vs. passive navigation. For Realism, there is a significant difference in Active vs. Passive conditions, with Passive scoring higher as indicated by two asterisks **, while Exo Active scored significantly lower than Exo Passive as marked by one asterisk *. For Possibility to Act, Active outperforms Passive with three asterisks ***, and Ego Active scores the highest among all combinations. Quality of Interface shows little difference between Ego and Exo, or Active and Passive, across all conditions. For Possibility to Examine, Passive scores slightly higher than Active with one asterisk *, and Ego Active scores highest overall. Self-Evaluation of Performance reveals significant differences, with Ego Active performing better than Exo Active and Passive conditions, indicated by one and three asterisks * and ***. Error bars represent standard deviations across conditions.}
\end{figure*}
The 19 PQ questions can be classified into 5 sub-categories: Realism (question 3, 4, 5, 6, 7, 10, 13), Possibility to Act (question 1, 2, 8, 9; Act for short), Quality of Interface (question 14, 17, 18, with scores reversed; Quality for short), Possibility to Examine (question 11, 12, 19; Examine for short) and Self-Evaluation of Performance (question 15, 16; Evaluation for short).

We ran a Shapiro-Wilk normality test on each sub-category, which showed that the PQ responses did not follow a normal distribution on any sub-category.
A follow-up Kolmogorov-Smirnov test showed that the data followed a Gamma distribution on both viewpoint and navigation factors.
Therefore, we used a GLMM model with a Log-Gamma link function.
Results showed the significance of \textit{Navigation} and $\textit{Viewpoint} \times \textit{Navigation}$ on 4 sub-categories (Realism, Act, Quality and Examine), and no significance of \textit{Viewpoint} and \textit{Group}.
A pairwise post-hoc Tukey's HSD test on $\textit{Viewpoint} \times \textit{Navigation}$ further showed that active has significantly higher realism, possibility to act and examine scores than passive, particularly in ego viewpoint. Specifically, ego$+$active received the highest average score on four sub-categories (Realism, Act, Examine and Evaluation). On Act and Evaluation, the post-hoc Tukey test showed that ego$+$active is significantly better than two other combinations. 

\textit{Perceived Workload}
\begin{figure*}[t]
\centering
  \includegraphics[width=\linewidth]{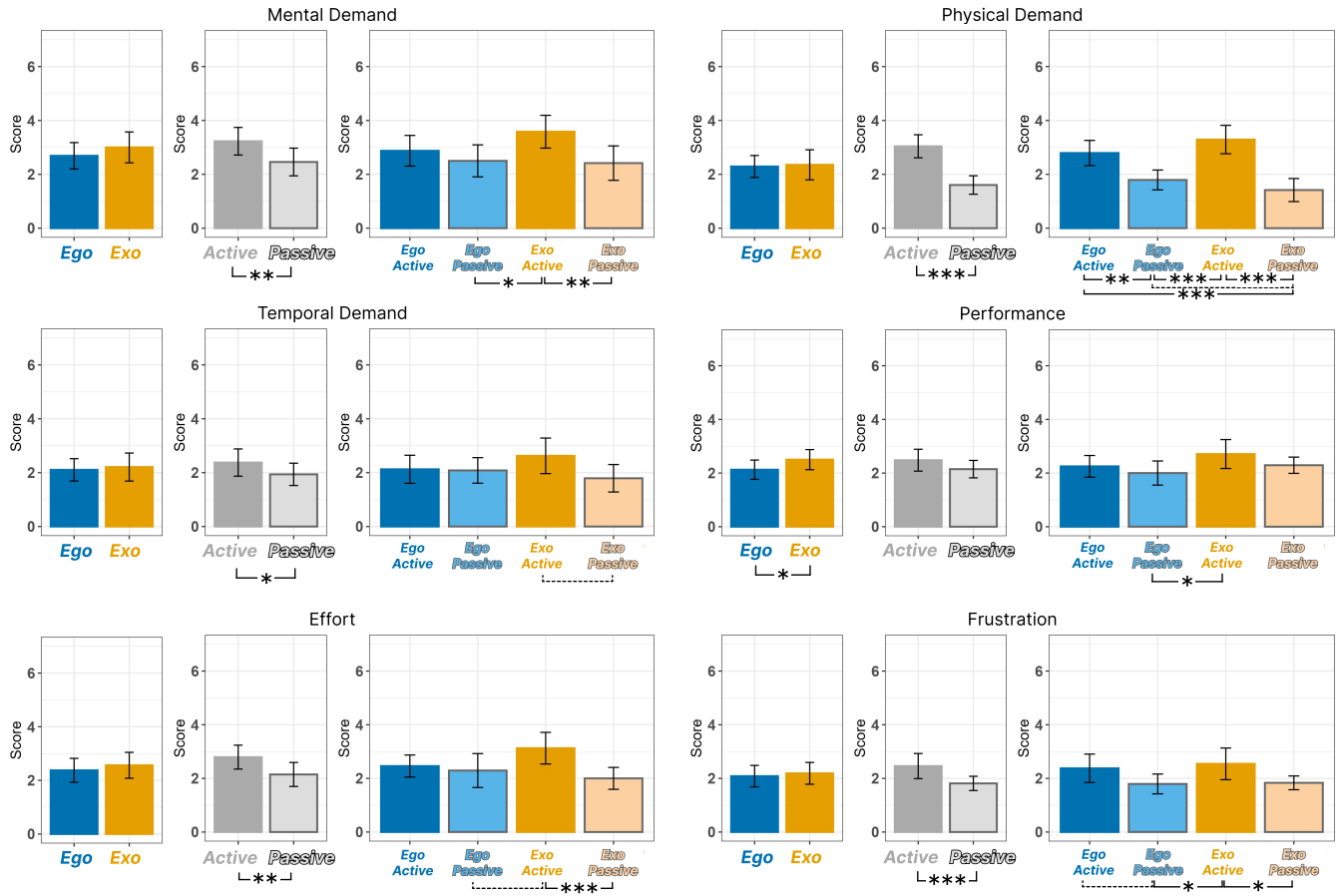}
  \caption{Average Sub-Scores of NASA-TLX Questionnaire. The dashed line indicates the marginal significance ($p < 0.1$) and the solid line indicates statistical significance ($p < 0.05$).}
  \label{fig:res_nasa}
  \Description{A series of bar charts displaying scores for Mental Demand, Physical Demand, Temporal Demand, Performance, Effort, and Frustration in a VR storytelling study, comparing egocentric (Ego) vs. exocentric (Exo) viewpoints and active vs. passive navigation. For Mental Demand, there is a significant difference between Active and Passive, with Passive scoring higher, indicated by two asterisks **, while Exo Active shows a significantly higher mental demand as marked by one and two asterisks * and **. Physical Demand reveals a substantial difference between Active and Passive, with Active scoring much higher, as shown by two and three asterisks ** and ***, particularly in the Ego Active condition. Temporal Demand shows a slight difference, with Passive scoring higher than Active, marked by one asterisk *. For Performance, Passive scores slightly higher than Active, with Exo Passive performing the best, marked by one asterisk *. Effort shows higher scores for Exo Active, with significant differences between conditions, marked by three asterisks ***. Finally, Frustration is significantly higher in Exo Active and Passive conditions, marked by one and three asterisks * and ***. Error bars represent standard deviations across the different conditions.}
\end{figure*}
We summarized the NASA-TLX reponses in \autoref{fig:res_nasa}. Expectedly, active had a higher overall workload than passive. 
We did the same step to test the data distribution and found that it also followed a Gamma distribution.
A GLMM with Log-Gamma link function comfirmed our observation with the signifcance on \textit{Navigation} ($p < 0.001$) and $\textit{Viewpoint} \times \textit{Navigation}$ ($p < 0.05$). It did not show any significance of other effects.
A follow-up post-hoc Tukey's HSD test on $\textit{Viewpoint} \times \textit{Navigation}$ further showed the significantly larger workload of exo$+$active than others, especially in physical demand.

{\textit{Result Summary.} Our hypothesis was supported by the results. The Ego+Active condition produced the strongest sense of presence without resulting in the highest perceived workload. However, we also observed that the heavy workload associated with active navigation might not be justified. In the Exo+Active condition, participants felt the least confident about their performance during story sessions. This suggests that the combined effects of viewpoint and navigation are more influential on story immersion than each factor individually.}

\subsection{Interview Analysis}
Finally, we collected and analyzed the final interviews about users' subjective experiences of reading stories in VR. Overall, We found a trend that \textbf{most participants preferred either ego$+$active or exo$+$passive stories and reported less favorable opinions towards the other two combinations.} We summarized the following reasons based on their comments.

\textbf{Active navigation provides freedom for explorations in the ego viewpoint.} Many participants favored active navigation primarily because of the ability to explore and play with the story elements. This is particularly true in two ego stories. For example, P4, P7, P21 and P23 all mentioned that the COVID story (Case 4) allowed them to freely move in the scene so that they could view the visualizations of airflows from different perspectives and understand the spread route of contaminants in the air. Also, P5, P6, P11 and P21 strongly endorsed the window open and fan installation operations as it gave them a sense of control over the story. Similiar comments appeared in the Tulsa story (Case 3) as well. Participants like the active navigation as \textit{``it is like an immersive exhibition in the museum''} (P4, P12, P20). This matches the highest PQ score of ego$+$active, as it greatly enhanced the diversity of interaction and range of movement in the story scene.

\textbf{Passive navigation is focused and efficient when reading stories in exo viewpoint.} While a considerable number of participants paid attention to user experiences, some prioritized the understanding and memorability of story content. Thus, they preferred the exo$+$passive condition because \textit{``information was concentrated in exo (viewpoint), and passive (navigation) had less workload than active (navigation)''} (P1, P9, P13, P14, P19). We observed that this group of participants did not have too much VR expertise, which increased their difficulty of active navigations and caused too much workload that impeded their focus (P19). With passive navigation, participants could quickly move to the next narratives without being trapped by VR interactions.

\textbf{Passive navigation resulted in overall negative user experiences in ego viewpoint.} The ego$+$passive condition received the most amount of critiques. Moving passively in an egocentric viewpoint (e.g. in a room) constrained the exploration participants could have. More seriously, it caused strong motionsickness for some participants (P1, P7). P7 even requested to pause the session and take off the VR headset for some rest. This indicates that our implementation of FOV reduction is not enough to alleviate motionsickness. Other techniques are necessary to be incorporated. Still, we received some positive feedback from participants who were less sensitive to movements in VR, who mentioned this experience was similiar to a Universal Studio tour on a club car (P4, P8). However, the overall current implementation of ego$+$passive is less acceptable to most participants.

\textbf{Active navigation increases the difficulty to keep track of spatial information in exo viewpoint.} As mentioned in feedback to exo$+$passive, active navigation introduced too much workload for VR non-experts in 2 exo stories. Another side effect we consistently received is the loss of object transform. This was particular the case for the 911 story (Case 1). P8, P9 and P23 reported that they could not tell WTC position as they rotated the map. This matches the lowest scores of exo$+$active in the spatial information depiction task.
\section{Key Findings and Discussion}
In this section, we summarize some key findings and compare our results with prior works to discuss the common and unique findings. We also discuss different expectations on immersive stories based on our particiapnts feedback.

\subsection{How does viewpoints influence story memorability?}
Our study shows that the viewpoint is a more decisive factor than the navigation in story memorability. 
\textbf{The exocentric viewpoint is slightly better than the egocentric one in the overview and detail memorization.} 
This result matches the disccovery of some prior works, in which the exocentric viewpoint benefits analytical tasks in immersive environments~\cite{yang2018maps, kraus2019impact, yang2020embodied}, but the advantage in our study is not as significant.
Our anticipated reason is that the exocentric viewpoint provide comprehensive information within audiences' view, which alleviates the problem of targeting story elements in space. However, this also causes a larger information density and more occlusions than the egocentric viewpoint. The relatively smaller scale of visual representations in exo viewpoints increases the attention to the story textual content received and prolongs the retention time in the memory so that they can did a better content comprehension. However, at the same time, audiences could become less sensitive to content transitions and therefore they might miss some important details. 
Overall, our results indicate that the exocentric viewpoint can be a better design if the primary goal is to promote the understanding and retention of narratives, and aforementioned problems should be considered to further improve the story.

In contrast, \textbf{the egocentric viewpoint demonstrated better memorability on spatial information than exocentric viewpoint.}
Similar result is also found in Krokos et al.'s~\cite{krokos2019virtual} and Yang et al.'s~\cite{yang2020virtual} work. 
\label{subsec:dis_view}
Objects in an egocentric viewpoin are often larger and more visible than the exocentric one. Resultedly, audiences can better recognize transforms (i.e. position, rotation and scale) and capture their changes.
More importantly, audiences are involved as part of a egocentric story. This enable them to use themselves as a frame of reference to infer transforms. In exocentric stories, they have to refer to one central element (e.g. the base map in \autoref{subsec:case1}) whose signficant transform change can greatly distract the spatial information perception since audiences need to find a new frame of reference every time.
Thus, the egocentric viewpoint is beneficial for those authors who want to convey understandings and perceptions related to the changes 3D spatial transforms (e.g. the renovation process of an office).

\subsection{How does navigation influence spatial immersion?}
As for the spatial immersion in our cases, navigation plays a more critical role than viewpoint. \textbf{Active navigation significantly enhances the sense of presence compared to passive, but their disparity of perceived workload varies between egocentric and exocentric conditions}. Specifically, we observed a substantial increase in workload for exocentric stories, while this effect was not as pronounced for egocentric ones.

This finding aligns with Lages et al.'s study~\cite{lages2018move}, which showed that walking-based navigation improved performance for users who lacked proficiency in VR manipulation. Our results suggest that the need for higher precision in finding specific angles in exocentric scenarios (e.g., \autoref{subsec:case1}, \autoref{subsec:case2}, and the grab-and-move tasks in Lages et al.'s study) may explain the increased workload. The ``changing frame of reference'' issue discussed in \autoref{subsec:dis_view} adds to the demand for spatial awareness and motor skills to achieve the correct angle, resulting in higher effort and frustration, as reflected in the study.

In contrast, egocentric stories primarily use teleportation for navigation, where the most challenging task is locating the target position. This task is not much more demanding than clicking a controller button for most participants. Consequently, we did not observe a significant increase in workload for active navigation in egocentric stories.

When designing navigation for VR stories, authors should consider the story's viewpoint based on these findings. However, focusing solely on individual navigation steps is insufficient. Ferguson et al.~\cite{ferguson2020role}, in their investigation of VR educational games, recommend that active navigation should seamlessly integrate with the story content and involve minimal learning curves. We argue that natural active navigation in VR storytelling should combine story progression with user exploration.

\subsection{What are audiences' expectations on VR stories?}
From our study, we identified a significant difference in audience priorities, which can be classified as either \textbf{information-oriented} or \textbf{experience-oriented}. This distinction greatly influences their expectations for VR stories.

Information-oriented audiences focus on discovering and understanding detailed content. They prefer full control over the information they consume and favor clear, structured layouts. This group tends to prefer author-driven experiences, where the narrative is pre-designed to ensure they do not miss key details. For these users, an exocentric viewpoint and passive navigation are ideal, as they offer greater information clarity and a well-organized, linear progression with minimal interaction.

On the other hand, experience-oriented audiences value a more user-driven experience. They seek control over VR elements and prefer embodied interactions, even if it means potentially missing parts of the story. These users often desire a personalized experience that supports multiple storylines for free exploration. An egocentric viewpoint and active navigation are better suited to this group, as they encourage deeper engagement. These audiences can be motivated to ``unlock'' different storylines through interactions, maximizing their freedom to explore the narrative.

{\subsection{Summary and Generalizability}}
{Our study underscores the significant impact of viewpoint and navigation on enhancing the memorability and spatial immersion of immersive stories. 
Among the tested conditions, Ego+Active and Exo+Passive were the most effective. The Ego+Active condition enhanced user immersion and creativity by allowing participants to actively explore and interact within the story, providing a strong sense of presence without significantly increasing the perceived workload. 
In contrast, the Exo+Passive condition improved content comprehension and recall by concentrating information and reducing cognitive load, making it ideal for users seeking a focused and efficient storytelling experience.}

{Immersive storytelling varies significantly across fields such as journalism, education, and entertainment, each with distinct objectives and audience expectations. 
In journalism, where clarity and factual accuracy are paramount, employing an ego viewpoint can enhance spatial understanding and engagement, making complex stories more accessible and memorable. 
In educational contexts, balancing ego and exo viewpoints can promote active learning and critical thinking by aligning storytelling approaches with educational goals. 
In entertainment, where immersion and emotional engagement are crucial, optimizing the combination of viewpoints and navigation styles can deepen the audience's connection to the narrative, enhancing their overall experience. 
Our preliminary exploration offers practical implications and guidance for making design decisions regarding viewpoint and navigation in immersive storytelling for these applications.}

{Additionally, future immersive stories could benefit from offering flexible viewpoints and navigation options to meet diverse audience needs for information and experiences, providing a personalized experience. 
For example, audiences might use an exocentric viewpoint to gain an overview and then ``zoom in" on areas of interest for more detailed information from an egocentric view. 
Similarly, active navigation can be advantageous for audiences proficient in VR who seek more engagement and immersion, while passive navigation provides a suitable way for those less familiar with VR or who prefer to focus solely on story content to control story progression.}

\section{Limitations and Future Work}
We adapted four selected web-based stories into immersive environments. 
To guide this adaptation, we conducted a formative user study to identify critical design considerations and explored various design options. Our exploration was not exhaustive; it aimed to facilitate the adaptation process by focusing on standard, off-the-shelf VR solutions. 
{This project concentrated on specific aspects of viewpoint and navigation: ego vs. exo and active vs. passive. While we believe these are unique and essential areas to investigate, we acknowledge a much broader design space for immersive storytelling beyond these aspects. 
For instance, more sophisticated VR locomotion techniques~\cite{di2021locomotion}, such as redirected walking, could enhance the active navigation experience.}
Similarly, advanced visual guidance methods~\cite{lange2020hivefive}, such as those for out-of-view scenarios~\cite{petford2019comparison, lin2021labeling} and multisensory approaches~\cite{melo2020multisensory}, offer additional potential for enhancing user experience. 
We did not incorporate these techniques, as our primary focus was the user study and providing easily accessible techniques available on standard platforms. 
Building a comprehensive design space to inform future immersive story design is crucial, and we consider this an area for future work.

Our study materials were adapted from web-based stories that were initially designed for web browsers and contained design patterns specifically tailored for web environments~\cite{bach2018narrative}. 
It is unclear whether these design patterns are still applicable in immersive environments or whether we need to develop unique and more appropriate design patterns for immersive stories. Studying the adaptation or innovation of these design patterns for immersive stories has the potential to bring about a more native immersive experience. 
{Furthermore, the web-based stories we adapted primarily follow a linear, single-scene structure. However, complex stories with multi-path or branching narratives also exist. 
While our studied scenario can serve as a foundational unit for these more complex stories, and our findings should be partially applicable, especially within their individual scenes. There is a need to explore transitions between scenes and more intricate active navigation, such as manipulating and selecting narrative branches.}

Through our study, we found that exocentric and egocentric viewpoints each have their own strengths and limitations. We considered viewpoint as an exclusive design element, testing it in a controlled manner. There is a great opportunity to design immersive stories that leverage both viewpoints to complement each other. In movies and data videos, cinematic techniques like camera movements and zooms have been used to switch focuses and direct audiences' attention~\cite{amini2015understanding, xu2022from, li2023geocamera}. Similar incoporation such as viewing some story pieces from an exocentric view and others from an egocentric view is also promising to investigate. Our study results can provide guidelines on which parts of the story to use which viewpoint. To achieve this, smooth transition techniques are necessary to allow seamless and meaningful switching between viewpoints, which we aim to explore in our next research phase.

Finally, one important aspect of active vs. passive interaction is the agency of the readers. In our study, readers were given the freedom to navigate by themselves with guidance. However, there are other types of freedom that could be provided to readers, such as allowing them to freely decide their focus or even rearrange the storyline. This may empower readers but also risks them getting lost, particularly due to the more flexible interactivity in immersive environments. Balancing exploration and narrative control is a fundamental question in storytelling~\cite{thudt2018exploration}. 
Studying the appropriate threshold or mechanism for this balance is an important research thread in immersive storytelling.

\section{Conclusion}
In this paper, we presented a user study investigating the effect of viewpoint and navigation on spatial immersion and understanding in immersive stories. To prepare the study materials, we first elicited design considerations from collected 3D spatial web stories. We then adapted four selected web stories to immersive environments based on these design considerations. Finally, we conducted a user study to empirically investigate the effect of viewpoints and navigation on spatial immersion and understanding.
Our results showed a marginal significance of viewpoints on story understanding and a strong significance of navigation on spatial immersion, with preferences for the Ego$+$Active and Exo$+$Passive cases. 
Our study indicates the associations between viewpoints and navigation and provides a preliminary exploration of their individual and combined effects on VR storytelling, which could benefit future immersive storytelling designs.



\bibliographystyle{ACM-Reference-Format}
\bibliography{reference}

\end{document}